\documentclass[lettersize,journal]{IEEEtran}
\usepackage{amsmath, amssymb, amsthm, amsfonts}
\usepackage[hidelinks]{hyperref}
\usepackage{graphicx}
\usepackage{booktabs}
\usepackage{bm}
\usepackage{cite}
\usepackage{color}
\usepackage{mathdots}
\usepackage{optidef}
\usepackage[caption=false,font=normalsize,labelfont=sf,textfont=sf]{subfig}

\usepackage{orcidlink}

\usepackage{accents}
\usepackage{cases}  
\usepackage{algorithm}      
\usepackage{algorithmic}      
\usepackage{eqparbox}
\usepackage{xargs}   
\usepackage{spconfpdkul}
\usepackage{todonotes}
\usepackage{pgfplots}   
\usepackage{ifthen}
\usepackage[acronym, shortcuts]{glossaries}  
\usepackage{makecell}

\newacronym{wasn}{WASN}{wireless acoustic sensor network}
\newacronym{danse}{DANSE}{distributed adaptive node-specific signal estimation}
\newacronym{rdanse}{R-DANSE}{robust distributed adaptive node-specific signal estimation}
\newacronym{rsdanse}{rS-DANSE}{relaxed simultaneous node-updating distributed adaptive node-specific signal estimation}
\newacronym{rsgevddanse}{rS-GEVD-DANSE}{GEVD-based relaxed simultaneous node-updating distributed adaptive node-specific signal estimation}
\newacronym{dansep}{TI-DANSE$^+$}{topology-independent distributed adaptive node-specific signal estimation PLUS[TO BE CHANGED]}
\newacronym{idanse}{iDANSE}{iterationless distributed adaptive node-specific signal estimation}
\newacronym{ti}{TI}{topology-independent}
\newacronym{tiidanse}{TI-iDANSE}{topology-independent iterationless distributed adaptive node-specific signal estimation}
\newacronym{idansep}{iDANSE$^+$}{iterationless distributed adaptive node-specific signal estimation PLUS[TO BE CHANGED]}
\newacronym{gevddansep}{GEVD-DANSE$^+$}{GEVD-based distributed adaptive node-specific signal estimation PLUS[TO BE CHANGED]}
\newacronym{gevd_or_dansep}{(GEVD-)DANSE$^+$}{(GEVD-based) distributed adaptive node-specific signal estimation PLUS[TO BE CHANGED]}
\newacronym{tdanse}{T-DANSE}{tree-DANSE}
\newacronym{tidanse}{TI-DANSE}{topology-independent distributed adaptive node-specific signal estimation}
\newacronym{tigevddanse}{TI-GEVD-DANSE}{topology-independent GEVD-based distributed adaptive node-specific signal estimation}
\newacronym{ti_or_gevddanse}{TI-(GEVD-)DANSE}{topology-independent (GEVD-based) distributed adaptive node-specific signal estimation}
\newacronym{gevddanse}{GEVD-DANSE}{GEVD-based distributed adaptive node-specific signal estimation}
\newacronym{gevd}{GEVD}{generalized eigenvalue decomposition}
\newacronym{gevc}{GEVC}{generalized eigenvector}
\newacronym{gevl}{GEVL}{generalized eigenvalue}
\newacronym{lmmse}{LMMSE}{linear minimum mean square error}
\newacronym{mmse}{MMSE}{minimum mean square error}
\newacronym{mwf}{MWF}{multichannel Wiener filter}
\newacronym{gevdmwf}{GEVD-MWF}{GEVD-based multichannel Wiener filter}
\newacronym{scm}{SCM}{spatial covariance matrix}
\newacronym{snr}{SNR}{signal-to-noise ratio}
\newacronym{mse}{MSE}{mean square error}
\newacronym{tad}{TAD}{target activity detector}
\newacronym{vad}{VAD}{voice activity detector}
\newacronym{spp}{SPP}{speech presence probability}
\newacronym{sad}{SAD}{source activity detector}
\newacronym{mst}{MST}{minimum spanning tree}
\newacronym{gsbcd}{GSBCD}{Gauss-Seidel block-coordinate descent}
\newacronym{ao}{AO}{alternating optimization}
\newacronym{mmut}{MMUT}{multiple max-$|\Uk^i|$ trees}
\newacronym{mc}{MC}{Monte-Carlo}
\newacronym{ac}{AC}{algebraic connectivity}
\newacronym{fc}{FC}{fully connected}
\newacronym{stft}{STFT}{short-time Fourier transform}
\newacronym{dft}{DFT}{discrete Fourier transform}
\newacronym{mil}{WMI}{Woodbury matrix identity}
\newacronym{poss}{POSS}{partially overlapping signal subspace}
\newacronym{foss}{FOSS}{fully overlapping signal subspace}
\newacronym{dof}{DoF}{degree of freedom}
\newacronym{dmwf}{dMWF}{distributed multichannel Wiener filter}
\newacronym{gevddmwf}{GEVD-dMWF}{GEVD-based distributed multichannel Wiener filter}
\newacronym{tidmwf}{TI-dMWF}{topology-independent distributed multichannel Wiener filter}
\newacronym{tigevddmwf}{TI-GEVD-dMWF}{topology-independent GEVD-based distributed multichannel Wiener filter}
\newacronym{tidmwfp}{TI-dMWF$^+$}{topology-independent distributed multichannel Wiener filter PLUS[TO BE CHANGED]}
\newacronym{ti_or_dmwf_or_p}{(TI-)dMWF($^+$)}{(topology-independent) distributed multichannel Wiener filter (PLUS[TO BE CHANGED])}
\newacronym{lcmv}{LCMV}{linearly constrained minimum variance}
\newacronym{gsc}{GSC}{generalized sidelobe canceller}
\newacronym{mvdr}{MVDR}{minimum variance distortionless response}
\newacronym{bf}{BF}{beamformer}
\newacronym{rtf}{RTF}{relative transfer function}
\newacronym{wlog}{w.l.o.g.}{without loss of generality}
\newacronym{stoi}{STOI}{short-term objective intelligibility}
\newacronym{gls}{GLS}{global-local subspaces}
\newacronym{fos}{FOS}{fully overlapping subspaces}
\newacronym{fods}{FODS}{fully overlapping desired subspaces}
\newacronym{pos}{PODS}{partially overlapping desired subspaces}
\newacronym{wola}{WOLA}{weighted overlap-add}
\newacronym{rir}{RIR}{room impulse response}
\newacronym{ser}{SER}{signal-to-error ratio}

\DeclareUnicodeCharacter{2212}{−}
\usepgfplotslibrary{groupplots,dateplot}
\usetikzlibrary{patterns,shapes.arrows,external}
\pgfplotsset{compat=newest}

\begin{document}

\title{Distributed Multichannel Wiener Filtering for Wireless Acoustic Sensor Networks}

\author{
	Paul Didier \orcidlink{0000-0002-1209-2614}, Toon van Waterschoot \orcidlink{0000-0002-6323-7350}, Simon Doclo \orcidlink{0000-0002-3392-2381}, Jörg Bitzer \orcidlink{0000-0002-0595-0262}, Pourya Behmandpoor \orcidlink{0000-0003-4522-1280}, Henri Gode \orcidlink{0009-0005-0764-7485}, and Marc Moonen \orcidlink{0000-0003-4461-0073}
	\thanks{This research work was carried out in the frame of Research Council KU Leuven project C14-21-0075 "A holistic approach to the design of integrated and distributed digital signal processing algorithms for audio and speech communication devices" and the European Union's Horizon 2020 research and innovation programme under the Marie Skłodowska-Curie Grant Agreement No. 956369: ``Service-Oriented Ubiquitous Network-Driven Sound — SOUNDS''. This paper reflects only the authors' views and the Union is not liable for any use that may be made of the contained information. The scientific responsibility is assumed by the authors.}
	\thanks{Paul Didier, Toon van Waterschoot, and Marc Moonen are with STADIUS Center for Dynamical Systems, Signal Processing, and Data Analytics, Department of Electrical Engineering (ESAT), KU Leuven, 3001 Leuven, Belgium (contact e-mail: \url{marc.moonen@kuleuven.be}).}
	\thanks{Simon Doclo and Henri Gode are with the Signal Processing Group, Department of Medical Physics, and the Acoustics and Cluster of Excellence Hearing4all, Carl von Ossietzky Universit{\"a}t Oldenburg, Oldenburg, Germany.}
	\thanks{Simon Doclo and Jörg Bitzer are with Fraunhofer IDMT, Project Group Hearing, Speech and Audio Technology, Oldenburg, Germany.} 
	\thanks{Pourya Behmandpoor is with the Department of Electronics and Informatics, Vrije Universiteit Brussel, B-1050 Brussels, Belgium.}
}

\maketitle

\glsunset{scm}
\glsunset{tidmwf}
\glsunset{tigevddmwf}
\glsunset{gevddmwf}
\glsunset{gevdmwf}
\glsunset{tidanse}
\glsunset{rsdanse}
\glsunset{rsgevddanse}
\glsunset{rdanse}

\newif\ifshrunk  
\shrunktrue  

\begin{abstract}
    In a wireless acoustic sensor network (WASN), devices (i.e., nodes) can collaborate through distributed algorithms to collectively perform audio signal processing tasks. This paper focuses on the distributed estimation of node-specific desired speech signals using network-wide Wiener filtering. The objective is to match the performance of a centralized system that would have access to all microphone signals, while reducing the communication bandwidth usage of the algorithm.
    Existing solutions, such as the distributed adaptive node-specific signal estimation (DANSE) algorithm, converge towards the multichannel Wiener filter (MWF) which solves a centralized linear minimum mean square error (LMMSE) signal estimation problem. However, they do so iteratively, which can be slow and impractical. Many solutions also assume that all nodes observe the same set of sources of interest, which is often not the case in practice.
	
    To overcome these limitations, we propose the distributed multichannel Wiener filter (dMWF) for fully connected WASNs. The dMWF is non-iterative and optimal even when nodes observe different sets of sources. In this algorithm, nodes exchange neighbor-pair-specific, low-dimensional (fused) signals estimating the contribution of sources observed by both nodes in the pair.
    We formally prove the optimality of dMWF and demonstrate its performance in simulated speech enhancement experiments. The proposed algorithm is shown to outperform DANSE in terms of objective metrics after short operation times, highlighting the benefit of its iterationless design.
\end{abstract}

\begin{IEEEkeywords}
  	Distributed signal processing, wireless acoustic sensor networks, multichannel Wiener filter, dimensionality reduction, distributed noise reduction
\end{IEEEkeywords}

\section{Introduction}\label{sec:intro}


\IEEEPARstart{M}{icrophone} arrays have been widely used in various applications to tackle audio signal processing tasks such as noise reduction, acoustic echo cancellation, or dereverberation. They have traditionally been deployed in \textit{centralized} settings where the recorded signals are aggregated and processed in a central processing unit, the so-called fusion center. 
However, \textit{distributed} settings recently garnered increasing attention due to the possibility to build \glspl*{wasn}, as more and more devices capable of recording, processing, and wirelessly transmitting audio data (e.g., laptops, smartphones, hearing aids, smart speakers) are available in our surroundings. Such networks may be described as wireless communication graphs where each \textit{node} is a device equipped with one or more sensors (microphones). Nodes in a \gls*{wasn} may be arbitrarily positioned, rendering \glspl*{wasn} flexible, scalable, and capable of covering wider spatial areas than centralized systems. These important advantages have led to the development of distributed audio signal processing algorithms. 

The goal of a distributed algorithm is to reach a performance that is as good as that of an equivalent centralized system, while accounting for the inherent challenges of distributed settings, e.g., limited communication bandwidth, synchronicity, or wireless link failures~\cite{bertrand_applications_2011}.
In this paper, we focus on node-specific distributed signal estimation through multichannel Wiener filtering, i.e., the challenge of estimating node-specific desired signals through a \gls*{mwf} (solution to a \gls*{lmmse} problem) applied to multiple noisy signals by letting devices in a \gls*{wasn} exchange signals and share the computation load.
We specifically aim at addressing the communication bandwidth challenge present in distributed settings and explicitly choose not to consider other (albeit equally important) challenges such as, e.g., synchronicity~\cite{schmalenstroeer_multi-stage_2017} or optimal sensor selection~\cite{zhangMicrophoneSubsetSelection2018a}.

Signal estimation often relies on filtering techniques such as, e.g., \textit{beamforming} via the \gls*{mvdr} beamformer or the \gls*{lcmv} beamformer. These filters minimizes output power while maintaining a certain response toward the target~\cite{vanveenBeamformingVersatileApproach1988}. Beamforming typically requires estimation of (relative) acoustic transfer functions, which is challenging in practice. An alternative technique is Wiener filtering via the \gls*{mwf}, a \gls*{lmmse} \textit{estimator} which is computed via spatial covariance matrices (\glspl*{scm}) estimated via the available signals. The \gls*{mwf} is the focus of the present paper.

Attempts to formulate a distributed \gls*{mwf} with limited communication bandwidth usage~\cite{doclo_reduced-bandwidth_2009} have led to the introduction of the \gls*{danse} algorithm~\cite{bertrandRobustDistributedNoise2009a,bertrand_distributed_2010,bertrand_distributed_2010-1}, which converges towards the equivalent \gls*{mwf} through iterations. In \gls*{danse}, nodes only transmit dimensionally reduced (i.e., fused) versions of their local sensor signals, thereby partly relieving the communication bandwidth usage.
The \gls*{danse} algorithm and its extensions all involve algorithmic iterations to compute the distributed filter and \textit{optimality}, i.e., matching the performance of the equivalent centralized \gls*{mwf}, is only met after upon convergence (typically several tens of iterations in a static acoustic scenario)~\cite{bertrand_distributed_2010}.
The number of iterations required to reach convergence generally increases with the total number of nodes.
As each iteration involves time-averaging over several time-frames to estimate second-order statistics, the final application must tolerate significant delays before a reliable estimate is available.
This significantly limits the relevance of such algorithms in practical applications, particularly in time-varying acoustic environments where quick adaptivity is paramount.
It should be noted that, although non-iterative distributed filters have been proposed~\cite{hiokaDistributedBlindSource2011,jiaDistributedMicrophoneArrays2006}, they generally rely on heuristic arguments and do not aim at solving a particular optimization problem.

Most distributed solutions to the signal estimation problem, including the \gls*{danse} algorithm and its extensions, assume that all sound sources of interest are observed by all nodes in the \gls*{wasn} to guarantee optimality upon convergence. In other words, they assume that the same low-dimensional signal subspace is spanned by the desired signals of all nodes. We refer to this as a \textit{\gls*{fods}} scenario. In any given acoustic environment, a sound source is only observed by a node if the signal it generates contributes to the sensor signals captured at that node. The case may arise where a particular source of interest to a node does not contribute {significantly} to the local sensor signals of another node\footnote{In practice, every source will always have \textit{some} contribution to the signals recorded at each sensor in the environment. However, in some cases, this contribution may be small, even below the noise floor (e.g., the microphone self-noise), or buried under undesired sources contributions.} -- the source may, e.g., be too far or obstructed from that node. In such case, the subspaces spanned by the desired signals of different nodes do not fully overlap, which is referred to as a \textit{\gls*{pos}} scenario, and optimality of the aforementioned algorithms can generally\footnote{
It can be noted that, if some nodes do not observe some source(s) but the number of signals exchanged between nodes is larger than the total number of sources in the environment, the \gls*{rdanse} algorithm from~\cite{bertrand_robust_2009} may be applied to still iteratively reach the centralized solution. The \gls*{rdanse} algorithm slightly modifies the \gls*{danse} algorithm from~\cite{bertrand_distributed_2010}, using fused signal channels from nodes that do observe the source(s) unobserved by the updating node to define the desired signal at the updating node, thereby artificially reinstating the \gls*{fods} condition. We note that, in practice, it may be undesired or impossible to modify the desired signals of some nodes~\cite{bertrandDistributedSignalEstimation2012a}.
} not be guaranteed~\cite{bertrand_distributed_2010}.
The \gls*{pos} scenario has been investigated in the context of \gls*{danse} in~\cite{bertrandDistributedSignalEstimation2012a}. However, the heuristic solution proposed there remains iterative and non-optimal.

\subsection{Contributions}

This paper introduces the \textit{\gls*{dmwf}}, a distributed \gls*{mse}-optimal estimator that can be applied in \gls*{fc} \glspl*{wasn} to optimally estimate node-specific desired signals in \gls*{pos} scenarios. The \gls*{dmwf} is optimal in the sense that it matches the performance of an equivalent centralized \gls*{mwf} in terms of \gls*{mse}. The \gls*{dmwf} is also shown to be suitable for application in dynamic environments, as it does not require iterations (i.e., only one set of second-order statistics must be estimated) to reach optimality. Its communication bandwidth requirements are moderate, as the \gls*{dmwf} only requires nodes to transmit \textit{fused} versions of their local sensor signals.

The key enabling principle of the \gls*{dmwf} is that it allows nodes to exchange node-pair-specific, low-dimensional estimates of the contribution of sources observed by both the node sending the estimate and any other node in the \gls*{wasn}.
In~\cite{didierOneShotDistributedNodeSpecific2024}, an algorithm based on a similar idea was introduced. However, its applicability was restricted to scenarios in which each source in the environment is observed either by a single node or by all nodes in the network. Furthermore, this algorithm assumed prior knowledge of the specific components present in the local microphone signals, which is not available in practice.

The \gls*{dmwf} proposed in this paper not only extends the applicability of the algorithm from~\cite{didierOneShotDistributedNodeSpecific2024} to arbitrary \gls*{pos} scenarios, where each source may be observed by any subset of nodes in the network, but also eliminates the need for prior knowledge of the components present in the local microphone signals. An optimality proof is provided, formally demonstrating that the \gls*{dmwf} solution is equivalent to the centralized solution in an \gls*{fc} \gls*{wasn}.
The performance of the \gls*{dmwf} is assessed in numerical experiments considering speech enhancement scenarios. Its advantages over the state-of-the-art \gls*{danse} algorithm are made visible in terms of convergence speed and objective metrics.

\subsection{Paper organization}

In~\secref{sec:prob_statement}, the signal model and centralized solutions are defined. In~\secref{sec:dmwf}, the \gls*{dmwf} is formulated for application in \gls*{fc} \glspl*{wasn} and its optimality is formally proven.
The communication bandwidth usage and computational complexity of the \gls*{dmwf} is analyzed and compared to that of \gls*{danse} in~\secref{sec:comm_bandwidth_usage}.
Numerical experiments applying the \gls*{dmwf} to speech enhancement problems are provided in~\secref{sec:res}. Finally, conclusions are formulated in~\secref{sec:conclusion}.

\vspace*{.5em}

\textit{Notation}: Italic letters denote scalars and lowercase boldface letters denote column vectors. Uppercase boldface letters denote matrices. The superscripts $\cdot^\T$ and $\cdot^\Her$ denote the transpose and Hermitian transpose operator, respectively. The symbols $\E[\cdot]$, $\|\cdot\|_2$, and $\|\cdot\|_\mathrm{F}$ denote the expected value operator, Euclidean norm, and Frobenius norm, respectively. 
The operator $\bd[{\mathbf{A}_1,\dots,\mathbf{A}_N}]$ creates a block diagonal matrix with the matrices $\mathbf{A}_1,\dots,\mathbf{A}_N$ on its main diagonal.
The symbol $|\mathcal{A}|$ denotes the number of elements in the set $\mathcal{A}$. 
The symbol $\backslash$ is used to denote set exclusion.
The symbol $\zer_{A\times B}$ denotes the $A\times B$ all-zeros matrix and $\I[A]$ denotes the $A\times A$ identity matrix.
The set of complex numbers is denoted by $\mathbb{C}$.

\section{Problem Statement and Centralized Solution}\label{sec:prob_statement}

\subsection{Signal Model}\label{sec:signal_model}

Consider $K$ nodes constituting a \gls*{wasn}, where each node $k\in\K\triangleq\{1,\dots,K\}$ has $M_k$ sensors. The total number of sensors in the \gls*{wasn} is denoted by $M\triangleq \sum_{k\in\K} M_k$. This network is deployed in an acoustic environment where $Q$ mutually uncorrelated sound sources are present. Among those, $\Qn$ are \textit{noise sources} (not of interest to any node) and $\Qd = Q-\Qn$ are \textit{speech sources} (of interest to the nodes that observe them). We assume that the noise sources are always active (i.e., they do not produce speech signals), while the speech sources exhibit an ON-OFF behavior. In practice, other types of sources may be present, such as music or transient noise, but they are not considered here to enable \gls*{vad}-based \gls*{scm} estimation, as detailed in the next section. If \gls*{scm} estimation were to be performed using a strategy that does not rely on a \gls*{vad} (e.g.,~learned time-frequency masks as in~\cite{furnonDNNBasedMaskEstimation2021b}), assumptions on the nature of the sound sources may be relaxed.

We consider a \gls*{pos} scenario, i.e., where any given source may not be observed by one or more nodes. A node may be unable to observe a source if the node is located too far from the source or acoustically isolated from the source. Every speech source that is observed by a node is a \textit{desired source} for that node. Node-specific source interests (where a speech source may be not be of interest to a node, even if it is observed by that node) may occur in practice~\cite{plata-chavesDistributedSignalEstimation2015} and the \gls*{dmwf} definition may be extended to include such cases, yet this is not considered here for the sake of a simpler exposition.

Each source generates a latent signal which is assumed to be complex-valued to allow frequency-domain representations, e.g., in the \gls*{stft} domain.
At time $t$, the $m$-th sensor of node $k$ collects the signal $y_{k,m}[t]$, which is one entry of the $M_k$-dimensional node-specific signal vector $\mathbf{y}_k[t] \triangleq [y_{k,1}[t],\dots,y_{k,M_k}[t]]^\T\in\C[M_k]$.
In the following, we apply our theory to finite time-frames of signals, assuming that all signals are short-term stationary and ergodic, such that signal statistics can be estimated through averaging over several time-frames.
The signal model at node $k$ is then:

\begin{align}\label{eq:local_signalmodel}
  \yk[k][t] = \overbrace{\Ak[k][t]\slat[{}][t]}^{\sk[k][t]} + \overbrace{\Bk[k][t]\nlat[{}][t]}^{\nk[k][t]} + \vk[k][t]\fa k\in\K,
\end{align}

\noindent
where $\slat[{}][t]\in\C[{\Qd}]$ and $\nlat[{}][t]\in\C[{\Qn}]$ contain the stacked latent speech and noise source signals, respectively, while $\Ak[k][t]\in\C[M_k][{\Qd}]$ and $\Bk[k][t]\in\C[M_k][{\Qn}]$ are the corresponding speech and noise steering matrix (i.e., representing the effect of sound propagation from source to sensor, which may also be time-varying in dynamic acoustic scenarios), respectively. The vector $\vk[k][t]$ represents the self-noise, uncorrelated across sensors. Note that diffuse noise (correlated across the closely spaced sensors of a same node but not across widely spaced nodes) is included in the term $\nk[k][t]$ and not in $\vk[k][t]$.
For conciseness, the index $[t]$ will be discarded unless explicitly mentioned.

Any all-zeros column in $\Ak$ or $\Bk$ corresponds to a source that is not observed by node $k$. We assume that the sources are sufficiently spatially separated in the acoustic environment (and that no particular symmetries exist between the sources and sensors of $k$) such that the non-zero columns of $\Ak$ and $\Bk$ are linearly independent. That is, 
the matrix obtained after removing the all-zeros columns from the compound steering $[\Ak\:|\:\Bk]$ is assumed to be full column rank\footnote{We here effectively assume that the number of sensors $M_k$ is larger than or equal to the number of sources observed by node $k$. As will be discussed in~\secref{subsec:dmwf_def_basis}, this assumption is not necessary to ensure the optimality of the proposed algorithm and is merely made at this stage for the sake of a simpler exposition. It will be further relaxed in~\secref{subsec:dmwf_def_basis}.}.

Note that~\eqref{eq:local_signalmodel} can correspond to the narrowband approximation of the time-domain convolutive mixing process as a multiplication per frequency bin~\cite{avargelMultiplicativeTransferFunction2007}, in which case all subsequent expressions are frequency-bin-specific.
It is important to note that the narrowband approximation is generally only valid when the relative impulse responses between different sensors are short compared to the \gls*{stft} window length~\cite{kowalskiNarrowbandApproximationWideband2010}.

The node-specific signal model from~\eqref{eq:local_signalmodel} can be made network-wide by stacking the local sensor signal vectors as $\yk[] \triangleq [\yk[1]^\T,\dots,\yk[K]^\T]^\T\in\C[M]$, obtaining:

\begin{equation}\label{eq:centr_signalmodel}
  \yk[] = \overbrace{\Ak[]\slat}^{\sk[]} + \overbrace{\Bk[]\nlat}^{\nk[]} + \vk[],
\end{equation}

\noindent
where the node-specific speech and noise steering matrices are stacked in $\Ak[]\in\C[M][{\Qd}]$ and $\Bk[]\in\C[M][{\Qn}]$, respectively, and $\vk[] \triangleq [\vk[1]^\T,\dots,\vk[K]^\T]^\T\in\C[M]$.

We consider a scenario where nodes are interested in estimating the contribution of their desired sources, i.e., the part of the observed signal that contains speech, in a certain number $D$ of their local sensor signals.
For the sake of a simpler exposition, we assume that $D$ is equal for all nodes, as extension to a case where $D$ is different for each node is straightforward.
\Gls*{wlog}, we assume the first $D$ channels are of interest and define the so-called \textit{desired signal} at node $k$ as:

\begin{equation}\label{eq:dk}
  \dk \triangleq \Ekk^\T\sk = \Ek^\T\sk[]\in\C[D]\fa k\in\K,
\end{equation}

\noindent
where $\Ekk=[\mathbf{I}_D\:|\:\zer]^\T\in\{0,1\}^{M_k\times D}$ extracts the first $D$ channels of $\sk$ and $\Ek\in\{0,1\}^{M\times D}$ extracts the same $\dk$ from $\sk[]$. The goal of node $k$ is to estimate $\dk$ from available (noisy) signals.
We emphasize here that $\dk$ may be a mixture of individual speech signals.

\subsection{Observability and Source Groupings}\label{sec:obs_and_groupings}

We number all sources (speech and noise) consecutively as $1, 2,\dots,\Qd,\Qd+1,\dots,Q$.
Let $\Osetk$ denote the set of indices of all sources observed by node $k$. We let $Q_k\triangleq|\Osetk|\fa k\in\K$. If source index $j\not\in\Osetk$, then the corresponding column of the corresponding steering matrix is all-zeros ($\Ak$ if speech source, $\Bk$ if noise source, as used in~\eqref{eq:local_signalmodel}).
The indices of sources observed by both node $k$ and another node $q$ are contained in the set $\Osetk[kq] \triangleq \Osetk\cap\Osetk[q]$. We let $\Osetk[kk] = \Osetk\fa k\in\K$ and $Q_{kq}\triangleq |\Osetk[kq]|\fa k\in\K\fa q\in\Kbq[k]$, where $\Kbq[k] \triangleq \K \setminus \{k\}$.

The different types of scenarios are summarized as follows:
\begin{itemize}
  \item \textbf{\gls*{pos}}: any source may or may not be observed by any node (i.e., no constraints on the $\mathcal{O}$-sets).
  \item \textbf{\gls*{fods}}: all nodes observe the same set of \textit{desired} sources.
\end{itemize}

\begin{figure}[!ht]
  \centering
  \includegraphics[width=.8\columnwidth,trim={0 0 0 0},clip=false]{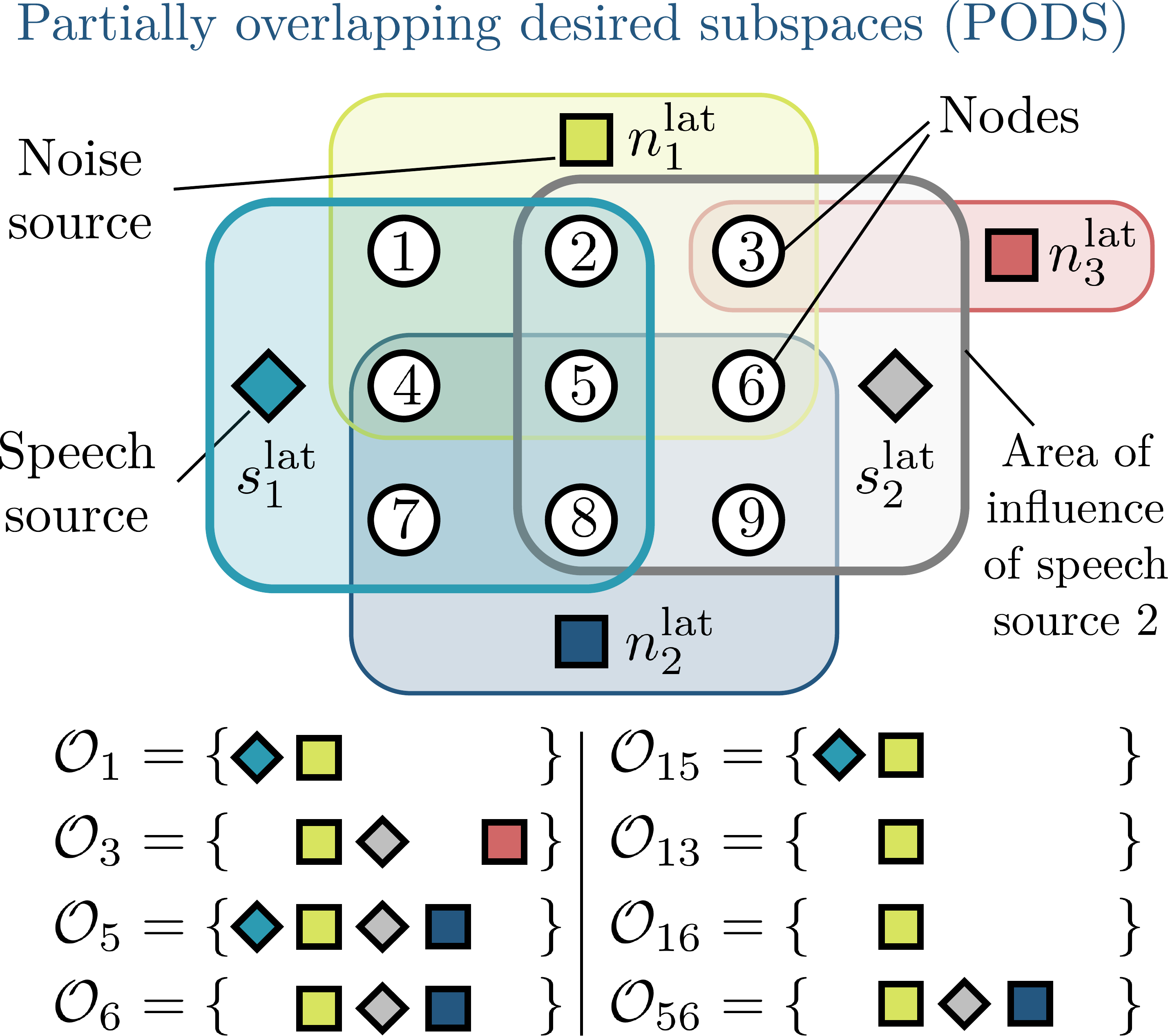}
  \caption{Example of \gls*{pos} scenario in a $K=9$ nodes \gls*{wasn}. Some sets of sources observed by a single node or a pair of nodes are shown as examples. Symbols are used instead of numbered indices for clear visualization, using diamonds ($\diamond$) for speech sources and squares ($\square$) for noise sources.}
  \label{fig:sigmod_pos}
\end{figure}

\begin{figure}[!ht]
  \centering
  \includegraphics[width=.8\columnwidth,trim={0 0 0 0},clip=false]{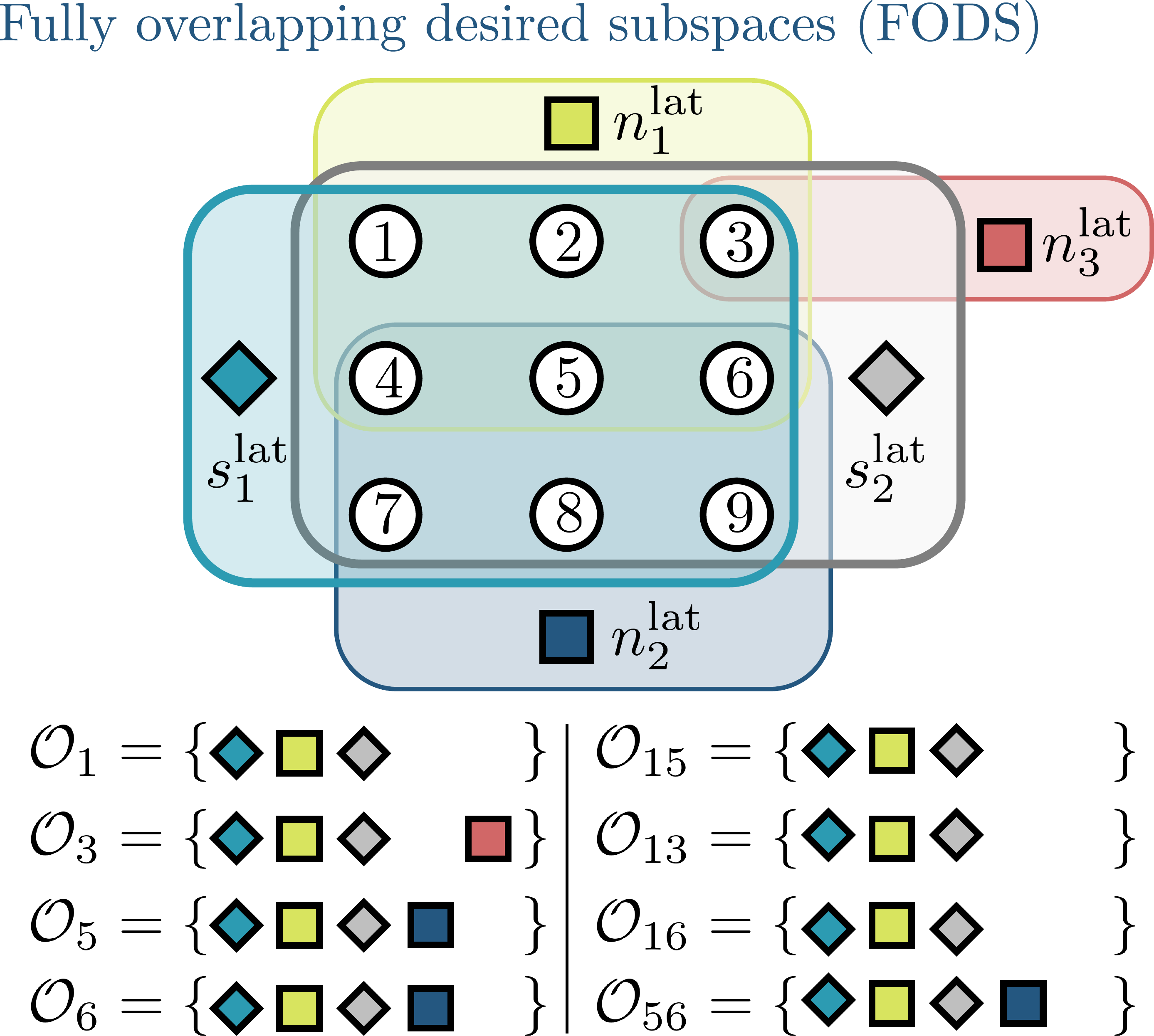}
  \caption{Example of \gls*{fods} scenario in a $K=9$ nodes \gls*{wasn}. Some sets of sources observed by a single node or a pair of nodes are shown as examples. The same symbols as in~\figref{fig:sigmod_pos} are used.}
  \label{fig:sigmod_fods}
\end{figure}

\noindent
Examples are provided in~\figref{fig:sigmod_pos} and~\figref{fig:sigmod_fods}.

\subsection{Centralized Signal Estimation}

Consider a centralized setting where the signals of all nodes in the \gls*{wasn} are aggregated at a fusion center. A general \gls*{pos} scenario is considered. In this case, the centralized signal vector $\yk[]$ can be used to estimate $\dk\fa k\in\K$. The following node-specific \gls*{lmmse} problem is defined:

\begin{equation}\label{eq:centralized_problem}
  \hWk \triangleq \argmin[{\mathbf{W}\in\C[M][D]}]
  \El[{
    \dk - \mathbf{W}^\Her\yk[]
  }],
\end{equation}

\noindent
which has the well-known \gls*{mwf} closed-form solution:

\begin{equation}\label{eq:centralized_mwf}
  \hWk = \Ryy^{-1}\Rydk,
\end{equation}

\noindent
with $\Ryy \triangleq \E[{\yk[]\yk[]^\Her}]$ and $\Rydk\triangleq\E[{
  \yk[]\dk^\Her
}]$. Note that, in general, $\Ryy$ will be full-rank and thus invertible due to the self-noise $\vk[]$.
In practice, $\dk$ is unavailable and $\Rydk$ must thus be indirectly estimated. It holds that:

\begin{align}\label{eq:Rydk_eq_RssEk_first}
  \Rydk &= \E[{
    \yk[]\dk^\Her
  }]
  =
  \E[{
    \left(
      \sk[] + \nk[] + \vk[]
    \right)\left(
      \Ek^\T\sk[]
    \right)^\Her
  }]\\
  &=
  \E[{
    \sk[]\sk[]^\Her
  }]\Ek
  \triangleq
  \Rss\Ek,\label{eq:Rydk_eq_RssEk}
\end{align}

\noindent
which shows that estimating $\Rss$ is sufficient to estimate $\Rydk$. 
Note that~\eqref{eq:Rydk_eq_RssEk} holds since speech sources are uncorrelated to noise sources and to self-noise. For the same reason, we can write $\Rss = \Ryy - \Rnn - \Rvv$. In speech enhancement scenarios, the \glspl*{scm} $\Ryy$ and $\Rnnv = \Rnn + \Rvv$ can be estimated based on $\yk[]$ by leveraging the ON-OFF structure of speech signals and the stationarity of typical noise signals.
By means of a \gls*{vad} or \gls*{spp} estimate, noise-only periods can be identified and used to estimate $\Rnnv$, while $\Ryy$ is estimated during speech+noise periods.
In acoustic scenarios including several speech sources ($\Qd>1$), $\dk$ is generally a mixture of individual speech signals and a multi-speaker \gls*{vad} or \gls*{spp} may be required. 
In such case, the estimation of $\Rnnv$ should be conducted when no speaker is active, while $\Ryy$ may be estimated when at least one speaker is active.
To enable the simple illustration of the \gls*{vad}-based \gls*{scm} estimation concept, we assume that all latent speech signals have the same ON-OFF behavior, implying that their activity can be monitored using a single \gls*{vad} or \gls*{spp}. Using exponential averaging with a forgetting factor $0\ll\beta< 1$ at time-frame $t$:

\begin{equation}\label{eq:expavg_centr}
  \begin{split}
    \Ryy[t] &= \beta\Ryy[t{-}1] + (1-\beta)\yk[]\yk[]^\Her\:\text{if speaker(s) active},\\
    \Rnnv[t] &= \beta\Rnnv[t{-}1] + (1-\beta)\yk[]\yk[]^\Her\:\text{if no speaker active}.
  \end{split}
\end{equation}

\noindent
Naturally, in practice, different speech signals will likely have different ON-OFF patterns. In such case, speaker-specific \glspl*{vad} or \glspl*{spp}~\cite{bertrandEnergybasedMultispeakerVoice2010,zhaoModelbasedDistributedNode2020} may be employed to identify, e.g., time-frames where no speaker is active (for $\Rnnv$ estimation) and time-frames where at least one speaker is active (for $\Ryy$ estimation).
Once $\Ryy$ and $\Rnnv$ are estimated, $\hWk$ can be computed via~\eqref{eq:centralized_mwf} and the centralized desired signal estimate at node $k$ is obtained as:

\begin{equation}
  \dhatk[k,\mathrm{centr}] \triangleq \hWk^\Her\yk[]\fa k\in\K.
\end{equation}

\section{Distributed MWF (dMWF)}\label{sec:dmwf}

In~\secref{sec:prob_statement}, the node-specific signal estimation solution was outlined for a centralized scenario.
In this section, we consider a \gls*{fc} \gls*{wasn} and define the \gls*{dmwf}, which allows each node to accurately estimate its desired signal while only transmitting low-dimensional fused versions of its local sensor signals to other nodes, without relying on a fusion center. 

\subsection{Preliminaries and Algorithm Structure}\label{subsec:dmwf_def_basis}

In a distributed signal estimation algorithm, an important challenge is the minimization of the amount of data that must be transmitted between nodes. This requirement is key to meet the communication bandwidth usage constraints of distributed systems, which are typically much more stringent than those of centralized systems~\cite{bertrand_applications_2011}.
A naive method to reach the same performance as an equivalent centralized system would be to let each node send all its local sensor signals to all other nodes. We will show in the following that this is not necessary, and that it is possible to reduce the amount of data transmitted while still reaching optimality.

Each node $k \in \K$ aims to estimate its desired signal $\dk$ using a \emph{network-wide Wiener filter} that incorporates all information available across the \gls*{wasn}.
Assume that node $k$ has access to (a subset or a linear combination of) the local sensor signals $\yk[q]$ from another node $q \in \Kbq[k]$. It may seem intuitive to assume that only (estimates of) signals from sources in the set $\Osetk[kq]$, i.e., the sources jointly observed by nodes $k$ and $q$, are relevant to node $k$. However, this intuition overlooks the potential benefit of using signals from sources that are not directly observed by node $k$: consider, for instance, a source $j$ that is observed by node $q$ but not by node $k$. The contribution of source $j$ to $\yk[q]$ may still be useful for network-wide Wiener filtering at node $k$ in multiple scenarios. For example, if there exists a third node $q' \in \K \setminus \{k, q\}$ that also observes source $j$, and which shares at least one source with node $k$, i.e., $\Osetk[kq'] \neq \emptyset$, then node $k$ can jointly exploit (a subset or a linear combination of) $\yk[q]$ and $\yk[q']$ to either (i) suppress interference from sources in $\Osetk[qq'] \setminus \Osetk[k]$, and/or (ii) enhance sources that are in $\Osetk[kq]$ or $\Osetk[kq']$.

Even if both $\Osetk[kq]$ and $\Osetk[kq']$ are empty, there may still exist a fourth node $q'' \in \K \setminus \{k, q, q'\}$ that observes source $j$ and satisfies $\Osetk[kq''] \neq \emptyset$. In this case as well, node $k$ can leverage $\yk[q]$ in conjunction with $\yk[q'']$ for estimating $\dk$ via network-wide Wiener filtering.
In general, the only source contributions that cannot be exploited by node $k$ are those originating from sources that are observed exclusively by a single other node.

We define the set $\Csetk$ (and its cardinality $\oQk$) of all sources observed by node $k$ \textit{and} at least one other node as:

\begin{align}\label{eq:obs_sources_sets}
  \Csetk &\triangleq \bigcup_{q\in\Kbq[k]}\Osetk[kq],\:\text{with}\:|\Csetk|\triangleq \oQk \:(\leq Q_k=|\Osetk|).
\end{align}

Now focusing on a node $q\in\K$, we can rewrite the signal model~\eqref{eq:local_signalmodel} to isolate the contribution of sources in $\Csetk[q]$:

\begin{align}
  \yk[q] &= \Ak[q]\slat + \Bk[q]\nlat + \vk[q]\\
  &= \left(
    \cAk[q] + \uAk[q]
  \right)\slat + \left(
    \cBk[q] + \uBk[q]
  \right)\nlat + \vk[q]\label{eq:sigmod_expanded_-1}\\
  &=
  \cAk[q]\slat + \cBk[q]\nlat + \uAk[q]\slat + \uBk[q]\nlat + \vk[q]\\
  &=
  \underbrace{\cCk[q]\xlat}_{
    \triangleq\cyk[q]
  } + \underbrace{\uCk[q]\xlat + \vk[q]}_{
    \triangleq\uyk[q]
  },\label{eq:sigmod_expanded}
\end{align}

\noindent
where $\cAk[q]$ and $\uAk[q]$ are equal to $\Ak[q]$ with the columns corresponding to sources outside of $\Csetk[q]$ and inside of $\Csetk[q]$ set to zero, respectively, and similarly for $\cBk[q]$ and $\uBk[q]$. 
We have also defined $\cCk[q] \triangleq [\cAk[q]\:|\:\cBk[q]]\in\C[M_q][Q]$, $\uCk[q] \triangleq [\uAk[q]\:|\:\uBk[q]]\in\C[M_q][Q]$, and $\xlat \triangleq [(\slat)^\T\:|\:(\nlat)^\T]^\T\in\C[Q]$ for conciseness.
The vectors $\cyk[q]$ and $\uyk[q]$ are the contributions of sources in $\Csetk[q]$ or outside of $\Csetk[q]$, respectively.

In order to reduce the amount of data transmitted through the \gls*{wasn}, the \gls*{dmwf} applies local sensor signals \textit{fusion}.
Any node $q\in\K$, instead of sending $\yk[q]$ to all other nodes, sends a low-dimensional (fused) linear combination of $\yk[q]$.
The fused signal sent from $q$ to other nodes must have a sufficient number of channels to entirely span the $\oQk[q]$-dimensional subspace to which $\cyk[q]$ belongs, such that:

\begin{equation}\label{eq:zktq}
  \zktq[q] \triangleq \Pktq[q]^\Her\yk[q]\in\C[{\oQk[q]}],
\end{equation}

\noindent
where $\Pktq[q]\in\C[M_q][{\oQk[q]}]$ is the so-called \textit{fusion matrix} at node $q$. The dimension of $\zktq[q]$, i.e., $\oQk[q]$, is specific to $q$ and depends on which sources node $q$ observes (cf.~\eqref{eq:obs_sources_sets}).
Naturally, if $M_q\leq \oQk[q]$, node $q$ can directly send all its local sensor signals to other nodes, bypassing the need for signal fusion. This trivial case is ignored from here on, and we assume $M_q> \oQk[q]\fa q\in\K$.
In practice, estimating $\oQk[q]$ may be done via source enumeration and labelling techniques~\cite{bahariDistributedLabellingAudio2016a,waxDetectionNumberSignals2021} providing knowledge of which sources are observed by node $q$ and which sources are observed by other nodes in the \gls*{wasn}.

\ifshrunk
\else
Note that the authors of~\cite{plata-chavesDistributedSignalEstimation2015} have derived a theoretical lower bound on the number of signals that must be exchanged between nodes of a \gls*{fc} \gls*{wasn} in order to achieve optimality, in scenarios where any given source may not be observed by (or may not be of interest to) one or more nodes. Importantly, this bound ensures optimality regardless of the noise observability patterns, i.e., regardless of whether each noise source is observed by any node or not.
Assuming that any speech source that is observed by a node is also of interest to that node, the lower bound as defined in~\cite{plata-chavesDistributedSignalEstimation2015} can be expressed as follows. In order for the distributed algorithm to achieve optimality under any noise observability pattern, each node $q\in\K$ should exchange observations of at least $L_q$ linearly independent signals, with:
\begin{equation}\label{eq:dmwf:lower_bound_pods}
  L_q = \min\{M_q,\mathrm{rank}\left(\unAk[-q]\right)\},
\end{equation}

\noindent
with $\unAk[-q] = \begin{bmatrix}
  \unAk[1],\dots,\unAk[q-1],\unAk[q+1],\dots,\unAk[K]
\end{bmatrix}\in\C[{\Qd}][D(K-1)]$, where $\unAk[n] = \Ekk[n]^\T\Ak[n]\in\C[{\Qd}][D]\fa n\in\K$.

The approach used to define the fused signal dimensions in the \gls*{dmwf} differs from~\cite{plata-chavesDistributedSignalEstimation2015}, as the noise observability patterns are here explicitly considered. Indeed, the set $\Csetk[q]$ (cf.~\eqref{eq:obs_sources_sets}), which is used to define the fused signals dimension $\oQk[q] = |\Csetk[q]|$ (cf.~\eqref{eq:zktq}), includes noise sources through the sets $\{\Osetk[qq']\}_{q'\in\Kbq}$.
This allows the number of fused signals transmitted by node $q$ in the \gls*{dmwf} to be potentially smaller than $L_q$ while retaining optimality (cf.~proof in~Appendix~1). This may occur if the acoustic environment contains, e.g., speech sources that are observed by nodes in $\Kbq$ but not by $q$. These sources contribute to the rank of $\unAk[-q]$ (hence to $L_q$) but not to $\oQk[q]$, possibly leading to $\oQk[q] < L_q$. We can also note that if the noise is uncorrelated across nodes (e.g., diffuse noise with large inter-node distances), the relationship $\oQk[q] \leq L_q\fa q\in\K$ is ensured. In general, $\oQk[q]$ may sometimes be greater than $L_q$, e.g., when noise sources observed by $q$ and at least one other nodes are present.
\fi

The definition of the fused signals $\{\zktq[q]\}_{q\in\K}$ in~\eqref{eq:zktq} allows the formulation of the \gls*{dmwf} as follows.
The \gls*{dmwf} is composed of two computing steps: (i) a \textit{discovery step} during which each node $q\in\K$ estimates $\Pktq[q]$ and (ii) an \textit{estimation step} during which each node $k\in\K$ uses the incoming fused signals to estimate $\dk$.
For clarity, we will first define the estimation step then focus on the discovery step, since an intuition for the latter can be formulated based on the former.

\subsection{Estimation Step}\label{subsec:dmwf_estimation_step}

With the fused signals defined in~\eqref{eq:zktq}, node $k\in\K$ has access to a so-called \textit{observation vector} $\ty$, defined as:

\begin{align}\label{eq:ty_dmwf}
  \ty &\triangleq 
  \begin{bmatrix}
    \yk^\T \:|\: \zktq[- k][k]^\T
  \end{bmatrix}\in\C[{\tMk}]\fa k\in\K\\
  \where
  \zktq[- k][k]^\T &\triangleq \begin{bmatrix}
    \zktq[1][k]^\T \:\dots\: \zktq[k-1][k]^\T \: \zktq[k+1][k]^\T\:\dots\: \zktq[K][k]^\T
  \end{bmatrix}^\T,
\end{align}

\noindent
and $\tMk \triangleq M_k + \sum_{q\in\Kbq[k]} \oQk[q]$.
Node $k$ can estimate its desired signal $\dk$ based on $\ty$ via:

\begin{align}\label{eq:estimation_lmmse_fc}
  \tW &\triangleq \argmin[{\mathbf{W}_k\in\C[{\tMk}][D]}]
  \El[{
    \dk - \mathbf{W}_k^\Her\ty
  }].
\end{align}

\noindent
The solution of~\eqref{eq:estimation_lmmse_fc} is the \gls*{dmwf} \textit{estimation filter}:

\begin{align}\label{eq:tW_dmwf_mwf}
  \boxed{\tW = \Rtyty^{-1}\Rtydk,}
\end{align}

\noindent
where $\Rtyty\triangleq\E[{
  \ty\ty^\Her
}]$ and $\Rtydk\triangleq\E[{
  \ty\dk^\Her
}]$.
The estimation filter $\tW$ can be partitioned as:

\begin{equation}\label{eq:tW_partitioning}
  \tW = \begin{bmatrix}
    \Wkk^\T,\: \Gkq[k][1]^\T \: \dots \: \Gkq[k][,k-1]^\T,\: \Gkq[k][,k+1]^\T \: \dots \: \Gkq[k][K]^\T
  \end{bmatrix}^\T,
\end{equation}

\noindent
where $\Wkk\in\C[M_k][D]$ is applied to $\yk$ and $\Gkq\in\C[{\oQk[q]}][D]$ is applied to $\zktq[q]\fa q\in\Kbq[k]$.
The final desired signal estimate at node $k$ is obtained as:

\begin{equation}\label{eq:dhatk_dmwf}
  \dhatk \triangleq \tW^\Her\ty = \Wkk^\Her\yk + \sum_{q\in\Kbq[k]}\Gkq[k][q]^\Her\zktq[q].
\end{equation}

As in the centralized case, the \glspl*{scm} in~\eqref{eq:tW_dmwf_mwf} may be estimated based on time-averaging.
Note that $\ty$ can be expressed as a function of the centralized signal vector, as $\ty = \Dk^\Her\yk[]$ where $\Dk\in\C[M][{\tMk}]$ contains an appropriate arrangement of the fusion matrices $\{\Pktq[q]\}_{q\in\Kbq[k]}$.
From~\eqref{eq:centr_signalmodel}, we can then write:

\begin{equation}\label{eq:ty_eq_ts_tn}
  \ty = \Dk^\Her\left(
    \sk[] + \nk[] + \vk[]
  \right)
  = \underbrace{\Dk^\Her\sk[]}_{\triangleq\ts} + \underbrace{\Dk^\Her\nk[] + \Dk^\Her\vk[]}_{\triangleq\tn}.
\end{equation}

\noindent
Under the assumption of mutually uncorrelated sources as for~\eqref{eq:Rydk_eq_RssEk}, it is then possible to expand $\Rtydk$ as:

\begin{equation}
  \Rtydk = \E[{\ts\ts^\Her}]\tEk=\left(
    \E[{\ty\ty^\Her}] - \E[{\tn\tn^\Her}]
  \right)\tEk,
\end{equation}

\noindent
where $\tEk\triangleq[\Ekk^\T\:|\:\zer]^\T\in\{0,1\}^{\tMk\times D}$. From there, a \gls*{vad}-based time-averaging strategy as in~\eqref{eq:expavg_centr} may for example be leveraged in speech enhancement scenarios to estimate $\Rtydk$ and compute $\tW$ via~\eqref{eq:tW_dmwf_mwf}.

\subsection{Discovery Step}\label{subsec:dmwf_discovery_step}

During the discovery step, the goal for node $q\in\K$ is to estimate $\oQk[q]$ channels of the contributions of sources observed by $q$ and at least one other node, i.e., $\cyk[q]$ as defined in~\eqref{eq:sigmod_expanded} (dimension $M_q \geq \oQk[q]$), to entirely span the $\oQk[q]$-dimensional subspace to which $\cyk[q]$ belongs.
An intuitive definition of the fusion matrix in~\eqref{eq:zktq} stems from an \gls*{lmmse} problem where node $q$ strives to estimate, \gls*{wlog}, the first $\oQk[q]$ channels of $\cyk[q]$ based on its local sensor signals $\yk[q]$. Denoting $\cdk[q] \triangleq \cEk[q]^\T\cyk[q]\in\C[{\oQk[q]}]$ the $\oQk[q]$-elements vector where $\cEk[q]\triangleq [\mathbf{I}_{\oQk[q]}\:|\:\zer]^\T\in\{0,1\}^{M_q\times \oQk[q]}$, we can write:

\begin{align}\label{eq:Pk_gkq_ideal}
  \Pk[q]&=\argmin[{\mathbf{P}_q\in\C[M_q][{\oQk[q]}]}]
  \El[{
    \cdk[q] - \mathbf{P}_q^\Her\yk[q]
  }]\\
  &=
  \Rykyk[q]^{-1}\Rykcyk[q]\cEk[q],\label{eq:Pk_gkq_ideal_mwf}
\end{align}

\noindent
where $\Rykcyk[q]\triangleq\E[{
  \yk[q]\cyk[q]^\Her
}]\in\C[M_q][{\oQk[q]}]$. In a hypothetical two-nodes scenario, if node $q$ could perfectly estimate $\cdk[q]$ and send it to node $k$, node $k$ would then have access to all the information needed to optimally estimate $\dk$ based on $\yk$ and $\zktq[q]$.
However, accurately estimating $\cdk[q]$ through~\eqref{eq:Pk_gkq_ideal} without access to the local sensor signals of all other nodes in the \gls*{wasn} may be challenging for node $q$. Indeed, not only can node $q$ only directly observe the full mixture $\yk[q] = \cyk[q] + \uyk[q]$, the sources that contribute to $\cyk[q]$ may be speech sources, noise sources, or (likely) a mixture of both, and the same applies to $\uyk[q]$. Since the sources contributing to $\uyk[q]$ cannot be guaranteed to have an ON-OFF behavior, \gls*{vad}-based estimation of $\Rykcyk[q]$ is generally not possible.

In order to overcome this (going back to the general case $K\geq 2$ nodes), we propose an alternative definition of $\Pk[q]$. First, we extract a subset of channels of $\yk$ and denote it as:

\begin{equation}\label{eq:yktq_subset}
  \bykq \triangleq \overbrace{[
    \I[{\oQk[q]}] \:|\: \zer
  ]}^{\Ektq^\T}\yk\in\C[{\oQk[q]}],
\end{equation}

\noindent
where $\Ektq\in\{0,1\}^{M_k\times \oQk[q]}$ is a selection matrix that selects the first $\oQk[q]$ channels\footnote{We will show in~\secref{subsec:dmwf_reduce_comm_bandwidth} that this number can be further reduced. We keep $\oQk[q]$ channels in this subsection for simplicity of exposition.} of $\yk$. As the notation suggests, $\bykq$ will be sent by node $k$ to node $q$ during the estimation step.
We then define the sum of all $\{\bykq\}_{k\in\Kbq}$ as:

\begin{equation}\label{eq:bykq_sum}
  \rsum[q] \triangleq \sum_{k\in\Kbq}\bykq\in\C[{\oQk[q]}].
\end{equation}

\noindent
From this, we formulate the alternative definition of $\Pk[q]$ as an \gls*{lmmse} problem where node $q$ strives to estimate $\rsum[q]$ based on its local sensor signals $\yk[q]$:

\begin{align}\label{eq:Pktq_alt}
  \Pktq[q] &= \argmin[{\mathbf{P}_q\in\C[M_q][{\oQk[q]}]}]
  \El[{
    \rsum[q] - \mathbf{P}_q^\Her\yk[q]
  }].
\end{align}

\noindent
The \gls*{lmmse} problem~\eqref{eq:Pktq_alt} has the \gls*{mwf} solution:

\begin{align}\label{eq:Pktq_alt_mwf}
  \boxed{
    \Pktq[q] = \Rykyk[q]^{-1}\Ryqrsum[q]\in\C[M_q][{\oQk[q]}]\fa q\in\K.
  }
\end{align}

\noindent
The fusion matrix definition~\eqref{eq:Pktq_alt} provides a practical approach for node $q$ to estimate $\Pktq[q]$ as it does not require direct access to $\cdk[q]$ as in~\eqref{eq:Pk_gkq_ideal}.
Indeed, during the \gls*{dmwf} discovery step, node $q$ can collect the signals $\{\bykq\}_{k\in\Kbq}$ sent by the other nodes in the \gls*{wasn} and build the sum $\rsum[q]$. This allows node $q$ to estimate $\Ryqrsum[q]$ by simply averaging $\yk[q]\rsum[q]^\Her$ over multiple time-frames.

Although~\eqref{eq:Pktq_alt_mwf} differs from the intuitive definition~\eqref{eq:Pk_gkq_ideal}, it still allows node $q$ to compute an optimal estimate of $\dk[q]$ during the \gls*{dmwf} estimation step via~\eqref{eq:tW_dmwf_mwf}. This optimality is justified by the fact that $\Ryqrsum[q]$ (in~\eqref{eq:Pktq_alt_mwf}) has the same column space as $\Rykcyk[q]$ (in~\eqref{eq:Pk_gkq_ideal_mwf}). Indeed, we can expand $\Ryqrsum[q]$ as follows:

\begin{align}
  \Ryqrsum[q] &\triangleq \E[{
    \yk[q]\rsum[q]^\Her
  }]\in\C[M_q][{\oQk[q]}]\\
  &=
  \sum_{k\in\Kbq}
  \E[{
    \yk[q]\yk^\Her
  }]\Ektq\\
  &=
  \sum_{k\in\Kbq}
  \E[{
    \cyk[q]\cyk^\Her
  }]\Ektq\\
  &= \sum_{k\in\Kbq}
  \cCk[q]\E[{
    \xlat(\xlat)^\Her
  }]\cCk^\Her\Ektq.\label{eq:Rykbykq_def_3}
\end{align}

\noindent
Equation~\eqref{eq:Rykbykq_def_3} holds since $\uyk[q]$ does not contain any contribution of sources in $\Csetk[q]$, and thus is uncorrelated to $\yk$, and vice-versa with $\uyk$ and $\yk[q]$. Since both $\cCk[q]$ and $\cCk$ contain all-zeros columns corresponding to sources outside of $\Csetk[q]$ and $\Csetk$, respectively, we can rewrite~\eqref{eq:Rykbykq_def_3} as:

\begin{align}\label{eq:Rykbykq_def_4}
  \Ryqrsum[q] &= \cCku[q]\sum_{k\in\Kbq}
  \E[{
    \cxlat(\cxlat[k])^\Her
  }]\cCku^\Her\Ektq,
\end{align}

\noindent
where $\cCku[q]\in\C[M_q][{\oQk[q]}]$ contains the columns of $\cCk[q]$ related to sources in $\Csetk[q]$ and $\cxlat\in\C[{\oQk[q]}]$ contains the corresponding latent signals.
The \gls*{scm} $\Rykcyk[q]$ introduced in~\eqref{eq:Pk_gkq_ideal} can be written in a similar fashion as:

\begin{align}
  \Rykcyk[q]
  &=
  \E[{
    \yk[q]\cyk[q]^\Her
  }]\cEk[q] =
  \E[{
    \cyk[q]\cyk[q]^\Her
  }]\cEk[q]\\
  &= \cCku[q]\E[{
    \cxlat(\cxlat)^\Her
  }]\cCku[q]^\Her\cEk[q].\label{eq:Rykcyk_exp}
\end{align}

\noindent
In both~\eqref{eq:Rykbykq_def_4} and~\eqref{eq:Rykcyk_exp}, $\cCku[q]$ is right-multiplied by a full-rank $\oQk[q]\times \oQk[q]$ matrix. Consequently, there exists a full-rank $\oQk[q]\times \oQk[q]$ matrix $\mathbf{X}_{kq}$ such that:

\begin{align}
  \Ryqrsum[q] &= \Rykcyk[q]\mathbf{X}_{kq}.
\end{align}

\noindent
This shows that $\Ryqrsum[q]$ in~\eqref{eq:Pktq_alt_mwf} has the same column space as $\Rykcyk[q]$ in~\eqref{eq:Pk_gkq_ideal_mwf}.

Using Lemma 1, it follows that the desired signal estimate $\dhatk$ from~\eqref{eq:dhatk_dmwf} is optimal when the fusion matrix $\Pk[q]$ is defined as in~\eqref{eq:Pktq_alt_mwf}$\fa q\in\K$.

\textbf{Lemma 1:} In the \gls*{dmwf}, if $\zktq[q]$ is linearly transformed by an arbitrary full-rank $\oQk[q]\times \oQk[q]$ matrix, the desired signal estimate $\dhatk$ from~\eqref{eq:dhatk_dmwf} is not affected.

\textit{Proof}:
Let $\Pktqm[q][k]\triangleq \Pktq[q]\mathbf{X}_q$ with $\mathbf{X}_q\in\C[{\oQk[q]}][{\oQk[q]}]$ a full-rank square matrix$\fa q\in\Kbq[k]$. Suppose now that $\Pktqm[q][k]$ is used instead of $\Pktq[q]$ in the \gls*{dmwf}$\fa q\in\Kbq[k]$. Then, the transformations $\{\mathbf{X}_q\}_{q\in\Kbq[k]}$ can be compensated for by the filters obtained when solving~\eqref{eq:estimation_lmmse_fc}. Indeed, with $\Gkqm = \mathbf{X}_q^{-1}\Gkq\fa q\in\Kbq[k]$:

\begin{align}
  \Gkqm^\Her\Pktqm[q][k]^\Her\yk[q]
  &= \Gkq^\Her\mathbf{X}_q^{-\Her}\mathbf{X}_q^\Her\Pktq[q]^\Her\yk[q] = \Gkq^\Her\zktq[q]\fa q\in\Kbq[k],
\end{align}

\noindent
and thus: $\Wkk^\Her\yk + \sum_{q\in\Kbq[k]}\Gkqm^\Her\Pktqm[q][k]^\Her\yk[q]
  = \tW^\Her\ty = \dhatk$.\hfill$\blacksquare$

\subsection{dMWF Overview and Optimality}\label{subsec:dmwf_optimality}

The discovery and estimation steps are both necessary for the functioning of the \gls*{dmwf} (computation of fusion matrices, then use of fused signals for estimating the desired signal). The estimation step must be performed by each node $k\in\K$ at every time-frame to continuously estimate the desired signal $\dk$. The computation of the fusion matrices $\{ \Pktq[q] \}_{q\in\K}$ via the discovery step can be done following different strategies.

One option is to allow a \textit{warm start} of the algorithm, where the discovery step is performed for a certain number of time-frames at the beginning of the operation of the \gls*{wasn}, after which only the estimation step is performed at every time-frame. This approach assumes that the acoustic environment remains relatively static after the initial discovery phase, e.g., in a meeting room with fixed speaker positions.
Alternatively, the discovery step can be performed \textit{periodically}, e.g., every $N_\mathrm{ds}$ time-frames, where $N_\mathrm{ds}$ is a design parameter that can be adjusted based on the dynamic properties of the environment. The more often the discovery step is performed, the better the \gls*{dmwf} can adapt to changes in the acoustic environment, at the cost of increased communication bandwidth usage.

\begin{figure}[h]
  \centering
  \includegraphics[width=.9\columnwidth,trim={0 0 0 0},clip=false]{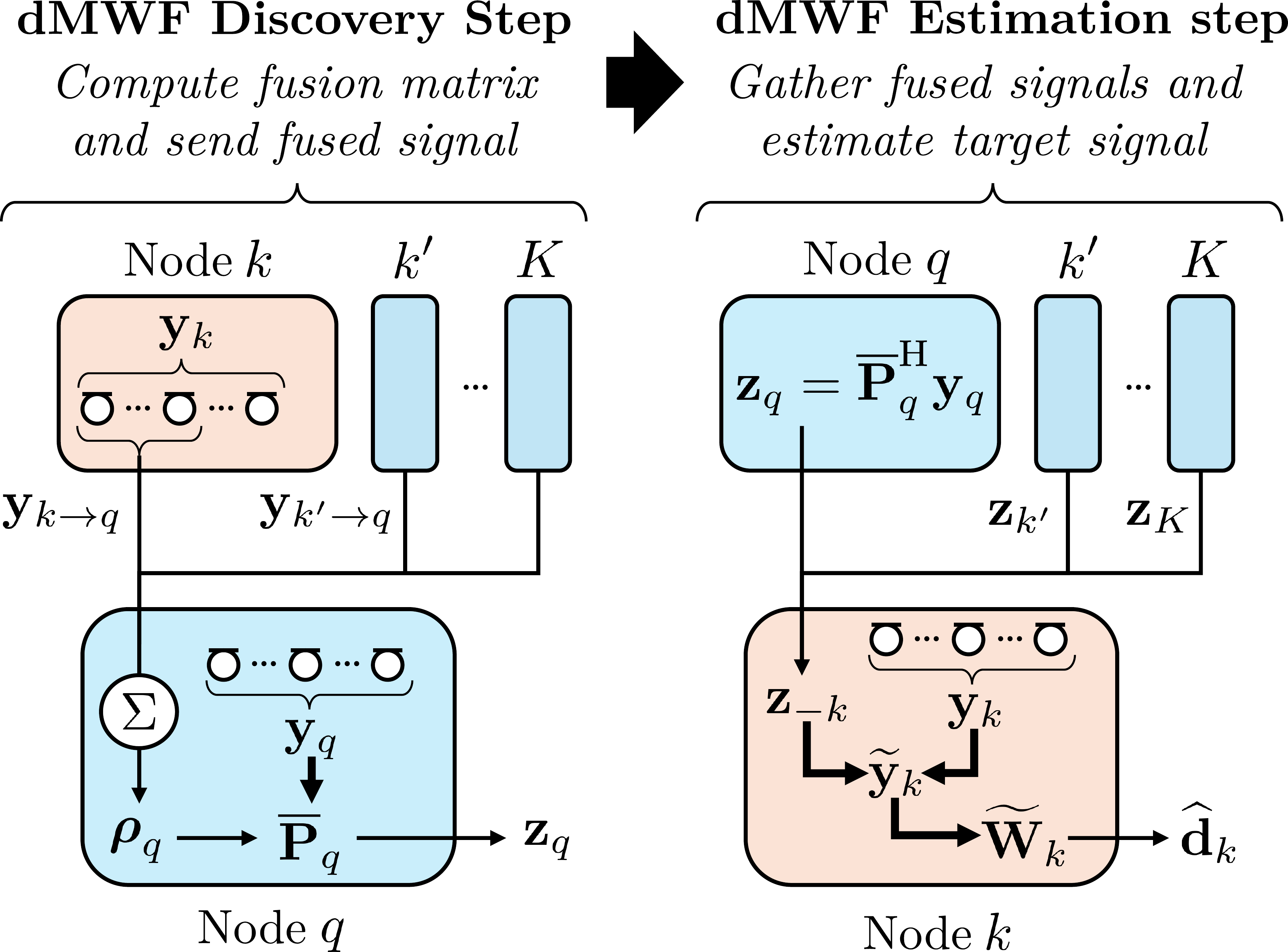}
  \caption{Schematic representation of the \gls*{dmwf}, focusing on node $q$ during the discovery step and node $k$ during the estimation step.}
  \label{fig:dmwf_schematic}
\end{figure}

The \gls*{dmwf} steps are summarized in~\figref{fig:dmwf_schematic} and~\algref{alg:dmwf} (for periodic fusion matrices updates). The optimality of the \gls*{dmwf} is formulated in Theorem~1.

\begin{algorithm}[h]
  \caption{The dMWF (with \textit{periodic} discovery step).}\label{alg:dmwf}
  \begin{algorithmic}[1]
    \FOR{time-frame index $t=0,1,\dots$}
      \IF{{
        $\mod(t,N_\mathrm{ds})=0$ (\textit{Discovery Step})
      }}
        \FOR{$k\in\K$}
            \STATE Node $k$ sends $\bykq[k][q][t]$ to node $q\fa q\in\Kbq[k]$.
        \ENDFOR
        \FOR{$q\in\K$}
          \STATE Node $q$ estimates $\Rykyk[q]$.
          \STATE Node $q$ computes $\rsum[q][t]$ as in~\eqref{eq:bykq_sum}.
          \STATE Node $q$ estimates $\Ryqrsum[q]$.
          \STATE Node $q$ updates $\Pktq[q]$ as in~\eqref{eq:Pktq_alt_mwf}.
        \ENDFOR
      \ENDIF
      \FOR{$q\in\K$}
        \STATE {
          Node $q$ computes $\zktq[q][t]$ as in~\eqref{eq:zktq} and sends it to all other nodes in $\Kbq$.
        }
      \ENDFOR
      \FOR{$k\in\K$}
        \STATE Node $k$ gathers $\{\zktq[q][t]\}_{q\in\Kbq[k]}$.
        \STATE Node $k$ builds $\ty[k][t]$ as in~\eqref{eq:ty_dmwf}.
        \STATE Node $k$ estimates $\Rtyty$ and $\Rtydk$.
        \STATE Node $k$ computes $\tW$ as in~\eqref{eq:tW_dmwf_mwf}.
        \STATE Node $k$ computes $\dhatk[k][t]$ as in~\eqref{eq:dhatk_dmwf}.
      \ENDFOR
    \ENDFOR
  \end{algorithmic}
\end{algorithm}

\textit{Theorem 1:} Under the signal model~\eqref{eq:local_signalmodel}, the \gls*{dmwf} desired signal estimate is equal to the centralized desired signal estimate at node $k\fa k\in\K$, when $\Pktq[q]$ is defined as in~\eqref{eq:Pktq_alt_mwf}$\fa q\in\K$, i.e., $\tW^\Her\ty = \hWk^\Her\yk[]\fa k\in\K$, with $\hWk$ defined as in~\eqref{eq:centralized_problem} and $\tW$ defined as in~\eqref{eq:estimation_lmmse_fc}.

\textit{Proof}: see Appendix A.

\subsection{Reducing Communication Bandwidth Usage}\label{subsec:dmwf_reduce_comm_bandwidth}

Consider a node pair $k\in\K$ and $q\in\Kbq[k]$. Only the sources observed by both nodes $k$ and $q$ are relevant for their local collaboration. Consequently, we can further reduce the number of channels exchanged between nodes $k$ and $q$ during the \gls*{dmwf} discovery step, down to $Q_{kq} = |\Osetk[kq]|\leq \oQk[q]$ channels. Node $k$ can simply send the following signal to node $q$:

\begin{equation}\label{eq:yktq_subset_red}
  \bykqred \triangleq \begin{bmatrix}
    \mathbf{I}_{Q_{kq}} \:|\: \zer
  \end{bmatrix}\yk\in\C[{Q_{kq}}].
\end{equation}

\noindent
This reduces the amount of signals exchanged during the discovery step from $\sum_{k\in\K} \sum_{q\in\Kbq[k]} \oQk[q]$ to $\sum_{k\in\K} \sum_{q\in\Kbq[k]} Q_{kq}$.
Estimating $Q_{kq}$ may also be done via source enumeration and labelling techniques~\cite{bahariDistributedLabellingAudio2016a,waxDetectionNumberSignals2021}.
However, the received signals in node $q$ cannot generally be assumed to all have the same dimension.
They thus cannot be summed up directly as in~\eqref{eq:bykq_sum}. Instead, node $q$ can extend $\{\bykqred\}_{k\in\Kbq}$ to a common dimension $\oQk[q]$ by padding them, yielding:

\begin{align}\label{eq:bykq_extended}
  \bykq^\star &\triangleq \begin{bmatrix}
    \mathbf{I}_{Q_{kq}}\\
    \padder[kq]
  \end{bmatrix}\bykqred\in\C[{\oQk[q]}],
\end{align}

\noindent
where $\padder[kq]$ is a $(\oQk[q]-Q_{kq})\times Q_{kq}$ padding transform matrix. Letting $\padder[kq]=\zer_{(\oQk[q]-Q_{kq})\times Q_{kq}}$ is an intuitive choice. However, in case where $\oQk[q] > \max_{k\in\Kbq}Q_{kq}$, this would lead to all-zeros entries in $\rsum[q]$ and hence all-zeros columns in $\Pk[q]$, which in turn would render the estimation step \gls*{lmmse} problem~\eqref{eq:estimation_lmmse_fc} ill-conditioned. It is thus preferable, in practice, to use a padding transform that is not all-zeros, e.g., all-ones or random vectors. The choice of padding transform has been verified to have a limited impact on the theoretical performance of the algorithm, according to extensive simulations in ideal settings (i.e., using oracle \glspl*{scm}). The rest of the discovery phase remains unchanged, simply replacing $\bykq$ with $\bykq^\star$ in~\eqref{eq:bykq_sum}.

\section{Comparison with the DANSE Algorithm}\label{sec:comm_bandwidth_usage}

\subsection{Applicability}\label{subsec:applicability}

The \gls*{dmwf} is optimal in any \gls*{pos} scenario (including \gls*{fods}) while \gls*{danse} is only optimal\footnote{{
  Note again that the \gls*{rdanse} from~\cite{bertrand_robust_2009} can be used to allow \gls*{danse}-like signal estimation in \gls*{pos} scenarios, at the cost of an intrusive change in the definition of the desired signal at the updating node, which may be undesired at some nodes in practice~\cite{bertrandDistributedSignalEstimation2012a}.}} in \gls*{fods} scenarios~\cite{bertrand_distributed_2010}. Additionally, the \gls*{dmwf} does not require iterations to reach optimality, while the \gls*{danse} algorithm does. In other words, the \gls*{dmwf} only requires estimating one set of \glspl*{scm} per node (i.e., $\Rykyk[q]$ and $\Ryqrsum[q]$ for node $q$, and $\Rtyty$ and $\Rtydk$ for node $k$) to compute the desired signal estimate at each time-frame, while \gls*{danse} requires estimating a set of \glspl*{scm} once per iteration until convergence is reached. In practice, \gls*{scm} estimation in any algorithm (including the \gls*{dmwf}) typically requires time-averaging over multiple time-frames\footnote{Data-driven methods may be alternatively used, for potentially faster estimation of accurate second-order statistics~\cite{furnonDNNBasedMaskEstimation2021b}.}, which introduces a certain delay in the adaptation of the \gls*{dmwf} to changes in the acoustic environment. In that sense, the \gls*{dmwf} with time-averaging-based \gls*{scm} estimation is not an {instantaneously responsive} algorithm, but rather an \textit{iterationless} algorithm. This delay can be expected to increase significantly in \gls*{danse}, an iterative algorithm, due to the required estimation of \glspl*{scm} over multiple iterations to reach the optimal solution.

\subsection{Communication Bandwidth Usage}\label{subsec:comm_bandwidth_usage}

We analyze the respective communication bandwidth requirements of the \gls*{dmwf} and \gls*{danse} assuming peer-to-peer communications for all signal exchanges and a \textit{periodic} \gls*{dmwf} discovery step as described in~\secref{subsec:dmwf_optimality}. 

{
  To perform a \gls*{dmwf} discovery step time-frame, the number of signal channels that must be exchanged is $N_\mathrm{dis} = \sum_{k\in\K} \sum_{q\in\Kbq[k]} Q_{kq}$ (cf.~\secref{subsec:dmwf_reduce_comm_bandwidth}). To perform a \gls*{dmwf} estimation step time-frame, this number becomes $N_\mathrm{est} = \sum_{k\in\K} \sum_{q\in\Kbq[k]} \oQk[q]$ (cf.~\secref{subsec:dmwf_estimation_step}). Since estimation steps are performed continuously, the total number of signal channels exchanged per time-frame when using the \gls*{dmwf} is at least $N_\mathrm{est}$ and at most $N_\mathrm{dis} + N_\mathrm{est}$, depending on how often the discovery step is performed.
  In a fusion-center-free scenario where every node would transmit all their raw local sensor signals to all other nodes, this number would be $\overline{M}= \sum_{k\in\K} \sum_{q\in\Kbq[k]} M_q=M(K-1)$.
  Assuming the \gls*{dmwf} discovery step is performed every $N_\mathrm{ds}$ time-frames, the \gls*{dmwf} compression factor becomes, on average through time, $\cf[\gls*{dmwf}] = \overline{M} / \left(N_\mathrm{est} + N_\mathrm{dis}/N_\mathrm{ds}\right)$.
}

In the \gls*{danse} algorithm, the number of signal channels exchanged during each iteration is $N_\mathrm{D} = \sum_{k\in\K} \sum_{q\in\Kbq[k]} \Qd = K(K-1)\Qd$~\cite{bertrand_distributed_2010}. The corresponding compression factor is thus $\cf[\gls*{danse}] = \overline{M} / N_\mathrm{D} = M/(K\Qd)$.
Considering the \gls*{dmwf} compression factor obtained during the estimation step, it holds that if $\sum_{k\in\K} \sum_{q\in\Kbq[k]} (\oQk[q] + Q_{kq}/N_\mathrm{ds}) \leq K(K-1)\Qd$, then $\cf[\gls*{dmwf}] \geq \cf[\gls*{danse}]$, and vice-versa. Therefore, depending on the values of $\{\oQk[q]\}_{q\in\K}$, $\{Q_{kq}\}_{(k,q)\in\K^2}$, and $N_\mathrm{ds}$, the \gls*{dmwf} may require more or less communication bandwidth usage than \gls*{danse} (or \gls*{rdanse}~\cite{bertrand_robust_2009}).

\ifshrunk
\else
{
  Note that both the \gls*{danse} algorithm and the \gls*{dmwf} estimation step may opt for using broadcasting for signal exchange, since any given node sends the same fused signal to all other nodes. Broadcasting may be preferable in practice when direct peer-to-peer communication is not feasible or when it simplifies the network communication protocol, although it requires more energy per node. Broadcasting may also be possible during the \gls*{dmwf} discovery step, if the reduction in communication bandwidth usage suggested in~\secref{subsec:dmwf_reduce_comm_bandwidth} is not a primary concern and $\bykq$ can be sent instead of $\bykqred$. 
  When broadcasting is used for all signal exchanges, the number of signal channels exchanged per time-frame is $N_\mathrm{D,bc} = K\Qd$ for \gls*{danse} and between $N_\mathrm{est,bc} = \sum_{k\in\K} \oQk$ and $N_\mathrm{est,bc} + N_\mathrm{dis,bc} = \sum_{k\in\K} \sum_{q\in\Kbq[k]} 2\oQk$ for the \gls*{dmwf}, depending on how often the discovery step is performed. This yields a compression factor of $\cf[\gls*{danse},bc] = M/(K\Qd)$ for \gls*{danse} and $\cf[\gls*{dmwf},bc] = M / \sum_{k\in\K} (\oQk + \oQk/N_\mathrm{ds})$ (on average over time) for the \gls*{dmwf}. Here too, depending on the values of $\{\oQk\}_{k\in\K}$ and $N_\mathrm{ds}$, the \gls*{dmwf} may require more or less communication bandwidth usage than \gls*{danse}. A quantitative evaluation of this statement is provided in~\secref{subsec:dmwf:res_commbw}.
}
\fi

\subsection{Computational Complexity}\label{subsec:complexity}

In the \gls*{dmwf}, the most computationally costly steps are the estimation of the \glspl*{scm} $\Rtyty$ and $\Rtydk$ and the inversion of $\Rtyty$ to compute $\tW$ in~\eqref{eq:tW_dmwf_mwf}. Indeed, the fusion matrix estimation~\eqref{eq:Pktq_alt_mwf} is computationally cheaper as it involves smaller matrices.
The cost of the \gls*{scm} inversion can be reduced using the Woodbury identity down to the cost of an \gls*{scm} estimation step, as explained in~\cite{bertrand_distributed_2010}. We obtain the \gls*{dmwf} computational complexity for node $k\in\K$:

\begin{align}
  \mathrm{CC}_{\text{\gls*{dmwf}},k} &\triangleq 
  O\Bigg(
    \Bigg(
      M_k + \sum_{q\in\Kbq[k]} \oQk[q]
    \Bigg)^2
  \Bigg).
\end{align}

\noindent
This complexity is quadratic in the total number of channels in the fused signals $\sum_{q\in\Kbq[k]} \oQk[q]$ and the number of local sensors $M_k$. As generally $\oQk[q] \geq \Qd\fa q\in\Kbq[k]$, the \gls*{dmwf} computational complexity is slightly larger than that of \gls*{danse}, which is $\mathrm{CC}_{\text{\gls*{danse}},k} \triangleq O(
  (
    M_k + (K-1)\Qd
  )^2
)$~\cite{bertrand_distributed_2010}. Both in the \gls*{dmwf} and in \gls*{danse}, the \gls*{scm} inversion required to compute new estimation filters (based on newly collected data) is performed regularly, notably to account for possible changes in the acoustic environment. However, as the \gls*{dmwf} does not require iterations to reach optimality while \gls*{danse} does (and the number of necessary iterations to reach optimality may be high~\cite{bertrand_distributed_2010}), the \gls*{dmwf} can be expected to effectively require a lower compound computational effort (i.e., fewer \gls*{scm} inversions) to reach any given performance level.

\section{Simulation Results}\label{sec:res}

In this section, we provide simulated validation of the proposed \gls*{dmwf}. We compare its performance to that of an equivalent centralized system (where all individual sensor signals in the \gls*{wasn} are available at a fusion center), and to those of two variants of the \gls*{danse} algorithm: \gls*{danse}~\cite{bertrand_distributed_2010} (using sequential node-updating) and \gls*{rsdanse}, where all nodes update their parameters at each algorithmic iteration~\cite{bertrand_distributed_2010-1}. The \gls*{rsdanse} algorithm is expected to converge faster than \gls*{danse} but still requires iterations, unlike the \gls*{dmwf}.

First, a set of batch-mode simulations are performed to demonstrate the theoretical optimality of the \gls*{dmwf}.
Second, online-mode speech enhancement simulations are conducted in a more realistic, time-varying acoustic environment, to demonstrate the practical performance of the \gls*{dmwf} in comparison to the \gls*{danse} algorithm. In both sets of simulations, both \gls*{fods} and \gls*{pos} scenarios are considered.

\subsection{Simulations with Oracle SCMs}\label{subsec:res_batchmode}

We first evaluate the performance of the considered algorithms in an ideal simulation setting, for both \gls*{fods} and \gls*{pos} scenarios. These simulations serve to establish the validity of the \gls*{dmwf} theoretical framework and to assess the convergence of \gls*{danse} and \gls*{rsdanse} in an idealized setting.

The idealized setting used in the current section only is defined as follows. A \gls*{fc} \gls*{wasn} composed of $K=6$ nodes, each with $M_k=5$ sensors, is deployed in an environment with $\Qd=2$ speech sources and $\Qn=2$ noise sources (note that, in this setting, $M_k\geq Q\fa k\in\K$ holds).
The steering matrices entries used in the signal model~\eqref{eq:local_signalmodel} are independently drawn from the standard normal distribution.
The aim of each node is to estimate its single-channel desired signal $d_k$ ($D=1$), i.e., the desired signal may be different for each node.

All \glspl*{scm} needed to compute the \glspl*{mwf} in~\eqref{eq:centralized_mwf},~\eqref{eq:tW_dmwf_mwf}, and~\eqref{eq:Pktq_alt_mwf} (and cf.~\cite{bertrand_distributed_2010,bertrand_distributed_2010-1} for \gls*{danse} and \gls*{rsdanse}, respectively) are computed directly using oracle knowledge of the true steering matrices (i.e., in this section, we do not perform time-averaging over frames of data, instead assuming oracle availability of the required \glspl*{scm}). Specifically, the centralized \gls*{scm} is computed as $\Ryy = \Rss + \Rnn + \Rvv= \Ak[]\Rsslat\Ak[]^\Her + \Bk[]\Rnnlat\Bk[]^\Her + \Rvv$, where $\Rsslat$ and $\Rnnlat$ are diagonal matrices containing the power of the latent speech and noise signals, respectively, on their diagonal, and $\Rvv$ is the self-noise \gls*{scm} (also diagonal, as self-noise is uncorrelated across sensors). Note that all \glspl*{scm} involved are complex-valued. The \glspl*{scm} required for the distributed algorithms are computed based on the centralized \glspl*{scm}. For instance, the \gls*{dmwf} estimation step \gls*{scm} $\Rtyty$ used in~\eqref{eq:tW_dmwf_mwf} is computed as:

\begin{equation}
  \Rtyty = \E[{\ty\ty^\Her}] = \E[{\Dk^\Her\yk[]\yk[]^\Her\Dk}] = \Dk^\Her\Ryy\Dk,
\end{equation}

\noindent
where $\Dk$ contains an appropriate arrangement of the fusion matrices $\{\Pktq[q]\}_{q\in\Kbq[k]}$ (as used in~\eqref{eq:ty_eq_ts_tn}). An analogous reasoning can be applied to obtain the other \glspl*{scm} needed in the \gls*{dmwf}, as well as for those needed at every iteration of \gls*{danse} and \gls*{rsdanse}. Full expressions are omitted for brevity.
Due to the use of oracle \glspl*{scm} in this section, no actual latent signals are necessary to compute the \glspl*{mwf}. Instead, the \glspl*{scm} are computed based on the latent speech and noise powers, which are set to 1 for each speech source and each noise source. The self-noise \gls*{scm} is set to 0.01 for each sensor.

To simulate observability patterns, we randomly generate observability matrices $\Obsd\in\{0,1\}^{K\times \Qd}$ and $\Obsn\in\{0,1\}^{K\times \Qn}$ with equal probability of 0 or 1 for each entry. If the $(k,s)$-th entry of $\Obsd$ is 1, then node $k$ observes source $s$, otherwise the corresponding column of the steering matrix $\Ak$ is set to zero, and similarly for $\Obsn$ and $\Bk$. For \gls*{pos} scenarios, we ensure that there exists at least one node that does not observe all speech sources. For \gls*{fods} scenarios, all nodes must observe all speech sources (noise sources can still be partially observed).

In \gls*{pos} scenarios, the number of speech sources observed can differ from node to node. The \gls*{danse} algorithm does not originally account for this and lets each node transmit $\Qd$-dimensional fused signals, regardless of the actual number of speech sources observed. This can lead to rank-deficient \glspl*{scm} and computational instabilities. We refer to this original formulation as \gls*{danse}$_{\Qd}$. We introduce a modified version of \gls*{danse}, referred to as \gls*{danse}$_{\Qd_k}$, where nodes only transmit as many fused signal channels as the number of speech sources they observe. The rest of the algorithm remains identical to the original from~\cite{bertrand_distributed_2010}. The same modification is applied to \gls*{rsdanse}, introducing \gls*{rsdanse}$_{\Qd_k}$, as opposed to the original \gls*{rsdanse}$_{\Qd}$ from~\cite{bertrand_distributed_2010-1}.
Again, we note that the \gls*{rdanse} algorithm from~\cite{bertrand_robust_2009} may be used to maintain convergence in \gls*{pos} scenarios. However, a comparison was not included in this section as, in practice, it may be undesired or impossible to modify the desired signals of some nodes~\cite{bertrandDistributedSignalEstimation2012a}.

\begin{figure}[h]
  \centering
  \includegraphics[width=\columnwidth,trim={0 0 0 0},clip=false]{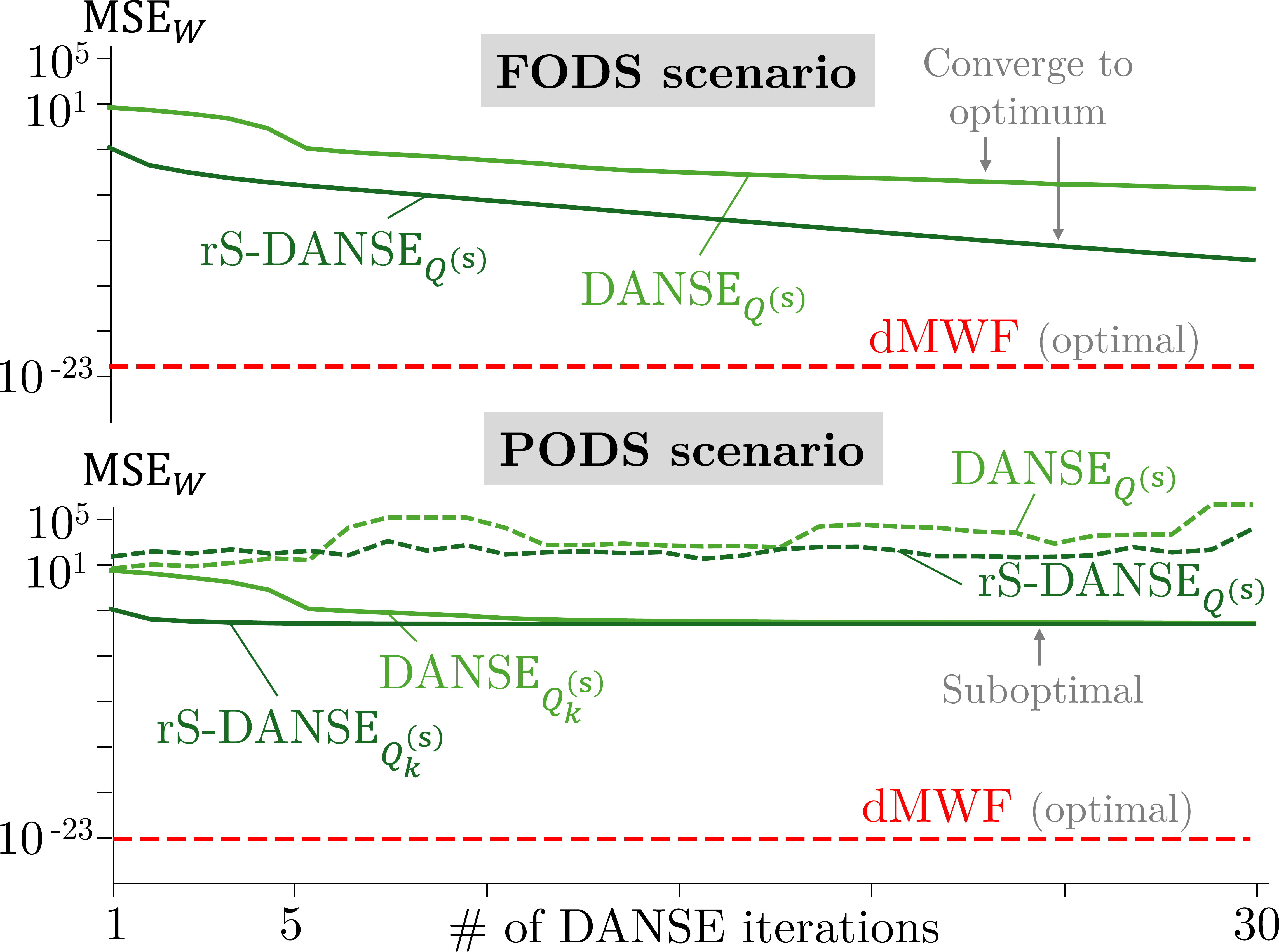}
  \caption{Average $\mathrm{MSE}_{W}$ over 10 randomly generated scenarios for the \gls*{dmwf}, \gls*{danse}, and \gls*{rsdanse}, in either \gls*{fods} (top) or \gls*{pos} scenarios (bottom).}
  \label{fig:res_batchmode}
\end{figure}

As evaluation metric, we consider the \gls*{mse} between the centralized \gls*{mwf} and the network-wide solutions\footnote{For full expressions of the network-wide filters, the reader is referred to~\eqref{eq:nw_dmwf} for the \gls*{dmwf} and (37) in~\cite{bertrand_distributed_2010} for \gls*{danse} and \gls*{rsdanse}.}, averaged over all nodes in the \gls*{wasn}:

\begin{equation}
  \mathrm{MSE}_{W} \triangleq \frac{1}{K}\sum_{k\in\K}\left\|\Wk - \hWk\right\|^2_\mathrm{F},
\end{equation}

\noindent
where $\hWk$ is the centralized \gls*{mwf} filter defined in~\eqref{eq:centralized_mwf}, and $\Wk$ is the network-wide filter at node $k$ of the algorithm under consideration (e.g.,~\eqref{eq:nw_dmwf} for \gls*{dmwf}). The \gls*{mse} obtained in this way for the \gls*{dmwf} and at the first 30 iterations of \gls*{danse} and \gls*{rsdanse}, in either \gls*{fods} or \gls*{pos} scenarios, are shown in~\figref{fig:res_batchmode} as averages over 10 randomly generated scenarios in each case.

The results confirm that the \gls*{dmwf} is optimal in both \gls*{fods} and \gls*{pos} scenarios, as it achieves an $\mathrm{MSE}_{W}$ down to numerical precision in both cases. The \gls*{danse} and \gls*{rsdanse} algorithms, on the other hand, only tend towards optimality in \gls*{fods} scenarios, which is expected from the theory~\cite{bertrand_distributed_2010,bertrand_distributed_2010-1} (with \gls*{rsdanse} converging faster than \gls*{danse}). In \gls*{pos} scenarios, \gls*{danse} and \gls*{rsdanse} do not reach optimality, which is consistent with the fact that they are not designed to perform optimally in such scenarios~\cite{bertrand_distributed_2010,bertrand_distributed_2010-1}. In particular, the ill-conditioning of the original formulations \gls*{danse}$_{\Qd}$ and \gls*{rsdanse}$_{\Qd}$ leads to numerical instabilities. The modified versions \gls*{danse}$_{\Qd_k}$ and \gls*{rsdanse}$_{\Qd_k}$ are more stable, but converge to suboptimal filters. The \gls*{dmwf} thus outperforms both \gls*{danse} and \gls*{rsdanse} when \gls*{scm} oracle knowledge is available. Naturally, in practice, the \gls*{dmwf} would not have access to such oracle knowledge, and these results serve to demonstrate the theoretical validity of the proposed algorithm. A more realistic scenario is considered in the next section.

\subsection{Simulations with Estimated SCMs in Dynamic Environment}\label{subsec:res_realistic}

\subsubsection{Setup}\label{subsec:res_setup}

We consider a \gls*{fc} \gls*{wasn} consisting of $K{=}6$ nodes with $M_1=1$, $M_2=2$, and $M_k{=}5\fa k\in\{3,4,5,6\}$. Its deployment is simulated in a 5\,m $\times$ 5\,m $\times$ 3\,m shoebox room with $\Qd{=}2$ speech sources and $\Qn{=}2$ noise sources. All walls have a 0.51 absorption coefficient. Inside the room, a $2.5$\,m $\times 3$\,m fully absorptive partition wall is added at $x=2.5$\,m. Source and nodes positions are randomly initialized in the room, with a minimum distance of 0.25\,m to walls and of 1\,m between nodes and sources. The sensor positions for each node are randomly initialized within a sphere of radius 5\,cm centered at the node position.
Communication links between nodes are assumed ideal.

We simulate 60\,s of signals, at a sampling rate of 16\,kHz. 
The signals are processed in the \gls*{wola} domain with a \gls*{dft} length of 1024 samples, a 50\% inter-frame overlap, and a square-root Hann window. Each frequency-bin is thus processed independently. 
The latent signals generated by the speech sources are taken from the VCTK database~\cite{veaux2017vctk}, the 
noise sources produce {speech-shaped} noise.
The latent noise signals are normalized in order to ensure a \gls*{snr} of 0\,dB at the reference sensor of node 1 in the initialized spatial configuration.
Self-noise is added at each sensor with a power equal to 1\% of the latent speech source power. Diffuse noise at -10\,dB \gls*{snr} with respect to the average latent speech signal power is added at each sensor~\cite{habets2008generating}.

\begin{figure}[h]
  \centering
  \includegraphics[width=\columnwidth,trim={0 0 0 0},clip=false]{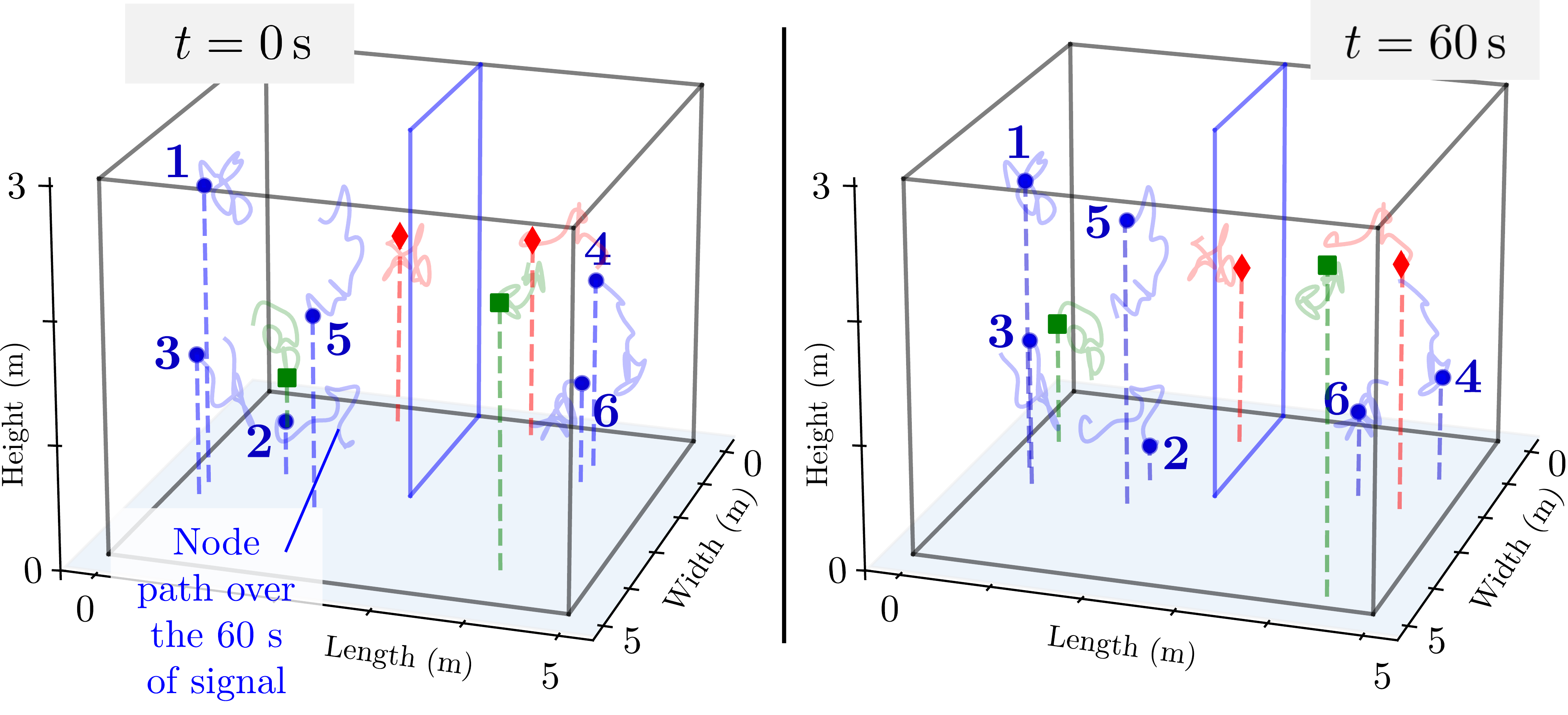}
  \caption{Example of a generated acoustic scenario. Initialization at $t=0$\,s (left) and layout after $t=60$\,s (right). The paths followed by nodes and sources across the 60\,s of simulated signal are shown by colored lines. The blue solid lines represent the edges of the absorptive partition wall.}
  \label{fig:dmwf:asc_example}
\end{figure}

The scenario is made \textit{dynamic} such that the nodes and sources positions randomly shift within a sphere of 20\,cm radius every 5\,s, always maintaining a minimum distance of 0.25\,m from the room walls and of 1\,m between nodes and sources. Within each 5\,s segment, nodes and sources remain static.
The \glspl*{rir} are generated within each static segment using the randomized image-source method~\cite{desenaModelingRectangularGeometries2015}.
An example of a randomly generated dynamic acoustic scenario is shown in~\figref{fig:dmwf:asc_example}.


In a real-world indoor scenario as simulated here, a source will never be entirely unobserved by any node, as some energy from the source will always be present at all sensors. However, the energy of a given source received at a given sensor might be small. This allows to define a threshold-based heuristic determining whether a source (speech or noise) is considered by the signal estimation algorithm as \textit{observed} by a node or not, based on received power at the reference sensor of that node.
We define the signal-to-interference ratio (SIR) at the reference sensor $m$ of node $k$ for source $s$ as the ratio (in dB) between the power of the received signal from source $s$ at $m$ and the power of the mixture of all other sources plus self-noise plus diffuse noise at $m$.
A source is then only considered observed by a node if its SIR for that source is at least $\delta$\,dB. Simulations are conducted for $\delta = 6$\,dB and $\delta = 10$\,dB, resulting in scenarios with more or less observed sources, respectively. The result is a so-called \textit{observability pattern}, which determines which nodes observe which sources. The observability patterns of the generated scenarios are kept fixed across the 60\,s of signals.

We perform a ``warm-start'' simulation where, at $t=0$\,s, all nodes are given access to a pre-computed batch estimate of the centralized \glspl*{scm} $\Ryy$ and $\Rnnv$ corresponding to the first static segment. This minimizes the impact of arbitrary \gls*{scm} initialization and allows comparison of the algorithms from a common starting point. From the start of the simulation on, a realistic exponential averaging strategy, as in~\eqref{eq:expavg_centr}, is applied to adaptively estimate the \glspl*{scm} needed for each algorithm in each frequency-bin with $\beta=0.967$ and an oracle energy-based \gls*{vad} computed on the clean speech signals.

As the frequency-bin-specific \glspl*{scm} are computed based on short-term data, the \glspl*{scm} can be ill-conditioned, especially in frequency-bins where the speech sources are weak. To improve numerical stability, the \gls*{mwf} used in each algorithm to estimate the desired signal is replaced by the \gls*{gevdmwf} defined in~\cite[Section III]{hassani_gevd-based_2016}, with rank $\Qd$ for all algorithms.

In the next section, the respective online performances of the local \gls*{gevdmwf}, the centralized \gls*{gevdmwf}, the \gls*{gevddanse} algorithm (where all fused signals are $\Qd$-dimensional)~\cite{hassani_gevd-based_2016}, its sequential node-updating version \gls*{rsgevddanse} (where all fused signals are $\Qd$-dimensional, obtained by applying \gls*{gevd}-based updates to \gls*{rsdanse}), and the \gls*{dmwf} where the estimation step update~\eqref{eq:tW_dmwf_mwf} is performed via a \gls*{gevdmwf}, are compared. The metrics obtained on the unprocessed (raw) first local sensor signals are also provided as a baseline.

For the \gls*{gevddanse} and \gls*{rsgevddanse} algorithms, the fusion matrices~\cite{bertrand_distributed_2010,bertrand_distributed_2010-1,hassani_gevd-based_2016} are updated every 20 time-frames (i.e., the algorithmic iteration index is incremented every 20 frames), while the \glspl*{scm} {and desired signal estimation filters} are updated using data from every time-frame (including time-frame from previous iterations), following the strategy proposed in~\cite{szurley_improved_2013}.
Over the 60\,s of signals, this setup results in 94 iterations for both \gls*{gevddanse} and \gls*{rsgevddanse}. In contrast, the \gls*{dmwf} continuously performs its estimation step at each time-frame and performs a discovery step every $N_\mathrm{ds}=8$ frames, cf.~\algref{alg:dmwf}.

\subsubsection{Speech Enhancement Performance}\label{subsec:res_online_results}

\begin{figure}[!ht]
  \centering
  \includegraphics[width=\columnwidth,trim={0 0 0 0},clip=false]{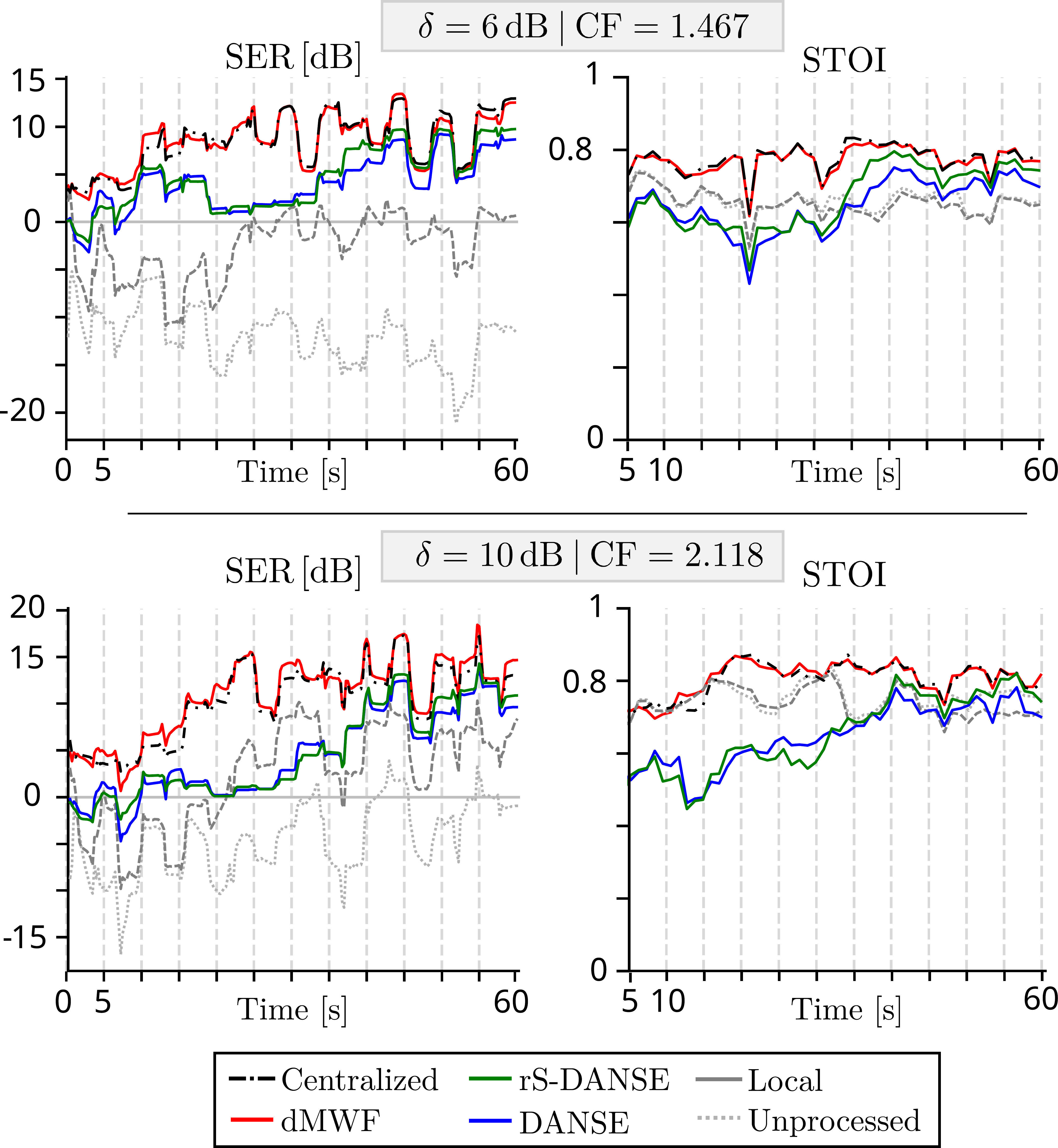}
  \caption{{Short-term \gls*{ser} (left) and \gls*{stoi} (right) at node 1, averaged across four randomly generated acoustic scenarios, for $\delta=6$\,dB (top) and $\delta=10$\,dB (bottom). The grey vertical lines indicate the times at which the sources and nodes move (cf.~\figref{fig:dmwf:asc_example}). The legend is shared across all plots.}}
  \label{fig:dmwf:res_online}
\end{figure}

The performance of each algorithm is evaluated in terms of \gls*{stoi}~\cite{taalAlgorithmIntelligibilityPrediction2011} and short-term \gls*{ser}, computed using the time-domain versions of the estimated desired signals $\widehat{d}_k[t]$ at each node $k\in\K$ and time $t$.
The \gls*{stoi} is computed every 1\,s for each observed speech source separately using the last 5\,s of signal{, taking into account only speech-active periods} (extracted via the oracle \gls*{vad}). The values are averaged over the observed speech sources for each node. The speech-active periods of time-domain desired signal $d_k[t]$ in each window are used as reference.
The \gls*{ser} is computed every 0.25\,s using the last 3\,s of signal, as:

\begin{align}
  \mathrm{SER}_k[t] &= 10\log_{10}\left(
    \frac{\sum_{\tau=t-\Delta}^{t}|d_k[\tau]|^2
    }{\sum_{\tau=t-\Delta}^{t}|d_k[\tau] - \widehat{d}_k[\tau]|^2}
  \right),
\end{align}

\noindent
where $t$ is the time-domain sample index at which the \gls*{ser} is computed (every 0.25\,s) and $\Delta=48000$ samples (corresponding to 3\,s).
The metrics obtained at node 1 are shown as averages over 4 randomly generated acoustic scenarios, in~\figref{fig:dmwf:res_online}. {Note that the \gls*{stoi} plot x-axis only begins at 5\,s, as the metric cannot be computed before that.}
In addition, the compression factor of the \gls*{gevddmwf} is computed as defined in~\secref{subsec:comm_bandwidth_usage} and averaged across the 4 randomly generated scenarios, yielding $\cf[\gls*{dmwf}] = 1.467$ for $\delta=6$\,dB and $\cf[\gls*{dmwf}] = 2.118$ for $\delta=10$\,dB. \gls*{gevddanse} and \gls*{rsgevddanse} reach a compression factor of 1.92, as all nodes transmit $\Qd=2$-dimensional fused signals.

The results show that the (\gls*{gevd}-based) \gls*{dmwf} outperforms both \gls*{gevddanse} and \gls*{rsgevddanse} in terms of both metrics, and in both $\delta=10$\,dB and $\delta=6$\,dB scenarios. The performance gap between the \gls*{dmwf} and the \gls*{danse} algorithms is more pronounced in the $\delta=6$\,dB scenario, where more sources are observed and nodes in the \gls*{dmwf} hence exchange more information. It can be noted that, even though the scenarios are understood as \gls*{pos} by the \gls*{dmwf} as it uses thresholds to determine observability, the scenario is \textit{effectively} \gls*{fods}, as all nodes observe all sources, albeit with different energy levels. This explains the relatively good performance reached by the \gls*{danse} algorithms after 40\,s, i.e., after about 10 iterations per node, as the \gls*{danse} algorithms are designed to perform optimally in \gls*{fods} scenarios. However, the \gls*{dmwf} still outperforms the \gls*{danse} algorithms practically, as it allows to quickly reach the centralized performance while efficiently tracking the time-varying scenario. The \gls*{danse} algorithms struggle to reach the centralized performance within the 60\,s of signals, even with the use of \gls*{gevd}-based updates and relatively fast fusion matrices updates (every 20 frames).

It is important to note that a careful definition of the observability of a source (e.g., here, through a choice of $\delta$ value), from the point of view of the \gls*{dmwf}, can allow the algorithm to reach a good performance while significantly reducing its communication bandwidth usage.
For instance, setting $\delta=10$\,dB, the \gls*{gevddmwf} achieves a compression factor of 2.118, which outbeats that of \gls*{gevddanse} (1.92) while still the algorithm still performs as well as the centralized \gls*{gevdmwf}.
\section{Conclusions}\label{sec:conclusion}

We introduced the \gls*{dmwf}, a distributed \gls*{mwf}-based algorithm for distributed signal estimation in a \gls*{fc} \gls*{wasn}.
In the \gls*{dmwf}, nodes estimate node-specific desired signals based on their local sensor signals and low-dimensional (i.e., fused) signals received from other nodes.
The \gls*{dmwf} is optimal in the sense that it yields the same solution as an equivalent centralized \gls*{mwf}, also in \gls*{pos} scenarios, i.e., where any source may or may not be observed by any node in the \gls*{wasn}.
The proposed algorithm provides clear advantages with respect to state-of-the-art solutions such as \gls*{danse}, which require multiple iterations to converge and only guarantee optimality in \gls*{fods} scenarios. 
Numerical experiments have demonstrated the practical relevance of the \gls*{dmwf} in realistic and time-varying acoustic scenarios.

The advantage of the \gls*{dmwf} (no reliance on iterative processes) may come at the cost of a slightly higher communication bandwidth usage than \gls*{danse}. However, this may be mitigated by the practical determination of source observability patterns. Our numerical experiments have indeed shown that the \gls*{dmwf} can outperform \gls*{danse} in terms of both performance and communication bandwidth usage upon careful selection of an observability threshold. 

\ifshrunk
\else
Possible avenues for future work include, but are not limited to, addressing the functioning of the \gls*{dmwf} in dynamic acoustic environments where the observability patterns change over time,
and formally evaluating the robustness of the \gls*{dmwf} to errors occurring when estimating the number of sources observed by each node.
\fi

\section*{Appendix A -- dMWF Optimality}\label{app:dmwf_opt}

In this section, we prove Theorem 1, i.e., the optimality of the \gls*{dmwf} under the signal model~\eqref{eq:local_signalmodel}.
The centralized \gls*{mwf} solution is first rewritten by making use of the \gls*{mil}.
The local fusion problem defined by the fusion matrix formulation in~\eqref{eq:Pktq_alt} is then also rewritten using the \glsreset{mil}\gls*{mil}. Based on these new formulations, we show that the fusion matrices that are used to send fused signals to node $k$ are in the same column space as the corresponding parts of the centralized filter at node $k\fa k\in\K$.
Using this observation, we show that the optimization conducted during the \gls*{dmwf} estimation step (cf.~\eqref{eq:estimation_lmmse_fc}) leads to the optimal centralized solution.

Based on~\eqref{eq:centr_signalmodel},~\eqref{eq:sigmod_expanded_-1}, and~\eqref{eq:sigmod_expanded}, we can write:

\begin{align}
    \yk[] 
    &= \overbrace{\left(
        \cAk[] + \uAk[]
    \right)\slat}^{=\sk[]} + \overbrace{\left(
        \cBk[] + \uBk[]
    \right)\nlat + \vk[]}^{=\nk[]}\label{eq:y_expanded_1}\\
    &=
    \underbrace{\cCk[]\xlat}_{\triangleq\cyk[]} + \underbrace{\uCk[]\xlat + \vk[]}_{\triangleq\uyk[]},\label{eq:y_expanded}
\end{align}

\noindent
where it holds that $\cyk[]=\begin{bmatrix}
    \cyk[1]^\T,\dots,\cyk[K]^\T
\end{bmatrix}^\T$, $\cAk[]\triangleq\begin{bmatrix}
    \cAk[1]^\T,\dots,\cAk[K]^\T
\end{bmatrix}^\T$, and similarly for $\uAk[]$, $\cBk[]$, $\uBk[]$, $\vk[]$, $\uyk[]$, $\cCk[]$, and $\uCk[]$. Using~\eqref{eq:y_expanded}, we can rewrite $\Ryy$ as:

\begin{align}
    \Ryy = \cCk[]\cRlat\cCk[]^\Her + \uRyy,\label{eq:Ryy_expanded_CRC}
\end{align}

\noindent
where $\cRlat\triangleq\E[{\xlat(\xlat)^\Her}]$ and $\uRyy\triangleq\E[{\uyk[]\uyk[]^\Her}]$. Note that, since $\uyk[]$ is entirely uncorrelated across nodes by definition, we have that $\uRyy = \bd[{\uRyy[1],\dots,\uRyy[K]}]$ where $\uRyy[q]\triangleq\E[{\uyk[q]\uyk[q]^\Her}]\fa q\in\K$. Since $\uyk[]$ includes $\vk[]$, which is uncorrelated across all sensors, we can assume that $\uRyy$ is invertible.
Applying the \gls*{mil} to $\Ryy$ gives:

\begin{align}\label{eq:invRyy_mil}
    \Ryy^{-1} =\Gam - \Gam\cCk[]\Xk[]^{-1}\cCk[]^\Her\Gam,
\end{align}

\noindent
where $\Gam \triangleq \uRyy^{-1}\triangleq\bd[{
    \Gam[1],\dots,\Gam[K]
}]$ (with $\Gam[q]=\uRyy[q]^{-1}\fa q\in\K$) and $\Xk[] \triangleq (\cRlat)^{-1} + \cCk[]^\Her\Gam\cCk[]\in\C[Q][Q]$. 
Considering that $\sk[] = (\cAk[] + \uAk[])\slat$ (cf.~\eqref{eq:y_expanded_1}), we can also expand $\Rss$ as:

\begin{align}
    \Rss &= \cAk[]\Rsslat\cAk[]^\Her + \uAk[]\Rsslat\uAk[]^\Her\\
    &=
    \cCk[]\Rsslatext\cCk[]^\Her + \uCk[]\Rsslatext\uCk[]^\Her,\label{eq:Rss_expanded_CRC}
\end{align}

\noindent
where
$\Rsslat \triangleq \E[{\slat(\slat)^\Her}]\in\C[\Qd][\Qd]$ and $\Rsslatext \triangleq \bd[{
    \Rsslat,\zer_{\Qn\times\Qn}
}]\in\C[Q][Q]$.
Combining~\eqref{eq:invRyy_mil} and~\eqref{eq:Rss_expanded_CRC} allows to rewrite the centralized \gls*{mwf} filter from~\eqref{eq:centralized_mwf} as:

\begin{equation}\label{eq:hWk_GamCM}
    \begin{split}
        \hWk  &=
        \left(\Gam - \Gam\cCk[]\Xk[]^{-1}\cCk[]^\Her\Gam\right)\times\\
        &\left(
            \cCk[]\Rsslatext\cCk[]^\Her + \uCk[]\Rsslatext\uCk[]^\Her
        \right)\Ek.
    \end{split}
\end{equation}

\noindent
Consider the partitioning $\hWk \triangleq \begin{bmatrix}
    \hWk[k1]^\T,\dots,\hWk[kK]^\T
\end{bmatrix}^\T$ where $\hWk[kq]\in\C[M_q][D]$ is the $q$-th submatrix of $\hWk$ applied to $\yk[q]$.
Expanding the right-most brackets of~\eqref{eq:hWk_GamCM}, we see that the term $\left(\Gam - \Gam\cCk[]\Xk[]^{-1}\cCk[]^\Her\Gam\right)\uCk[]\Rsslatext\uCk[]^\Her\Ek$ only affects $\hWk[kk]$, since $\uCk[]\Rsslatext\uCk[]^\Her$ is block-diagonal and $\Ek$ only extracts the rows corresponding to node $k$.

Consequently, with the partitioning $\cCk[]\triangleq\begin{bmatrix}
    \cCk[1]^\T,\dots,\cCk[K]^\T
\end{bmatrix}^\T$, we can write:

\begin{equation}\label{eq:hWk_GamCMk}
    \hWk[kq] =
    \Gam[q]\cCk[q]\Mk\fa q\in\Kbq[k],
\end{equation}

\noindent
where $\Mk \triangleq \left(
    \mathbf{I}_Q - \Xk[]^{-1}\cCk[]^\Her\Gam\cCk[]
\right)\Rsslatext\cCk[]^\Her\Ek\in\C[Q][D]$.
We can now use the fact that, by definition of the set $\Csetk[q]$ (cf.~\eqref{eq:obs_sources_sets}), only the columns of $\cCk[q]$ at indices corresponding to sources that are observed by node $q$ and at least one other node (i.e., the columns at indices $\Csetk[q]$) are non-zero.
Letting the full-column-rank matrix $\cCkqu[q]\in\C[M_q][{\oQk[q]}]$ denote $\cCk[q]$ with only the columns at indices $\Csetk[q]$ kept and $\Mkqu\in\C[{\oQk[q]}][D]$ denote $\Mk$ with only the corresponding rows kept, we can then write:

\begin{equation}\label{eq:hWk_GamCMk_red}
    \hWk[kq] =
    \Gam[q]\cCkqu[q]\Mkqu\fa q\in\Kbq[k].
\end{equation}

Consider now the \gls*{dmwf} local fusion problem~\eqref{eq:Pktq_alt}, where node $q$ estimates $\rsum[q]$ based on $\yk[q]$. The above reasoning can be repeated for this problem, replacing $\dk$ by $\rsum[q]$ and $\yk[]$ by $\yk[q]$. The \gls*{dmwf} fusion matrix at node $q$ can then be expressed as:

\begin{align}
    \Pktq[q] &=
    \Gam[q]\cCkqu[q]\Mqtk[qq],\label{eq:Pktq_GamCM_red}
\end{align}

\noindent
where $\Mqtk[qq]\in\C[{\oQk[q]}][{\oQk[q]}]$ is defined in an analogous way as $\Mkqu$, but using $\yk[q]$ instead of $\yk[]$ and aiming at estimating $\rsum[q]$ instead of $\dk$.
The full expression is omitted for the sake of brevity.
Due to the assumptions of source uncorrelatedness and steering matrices rank made in~\secref{sec:signal_model}, $\cRlat$ and $\cCkqu[q]$ are full-rank and full-column-rank, respectively. It can hence be shown that $\Mqtk[qq]$ is full-rank and, since it is square, also invertible.

A parallel can be made between~\eqref{eq:Pktq_GamCM_red} and~\eqref{eq:hWk_GamCMk_red}. The column space of $\Gam[q]\cCkqu[q][k]$ is indeed either transformed by $\Mkqu$ to lead to $\hWk[kq]$, or transformed by $\Mqtk[qq]$ to lead to $\Pktq[q][k]$. We will now show that the estimation phase of the \gls*{dmwf} yields the centralized solution.

Consider the partitioning of the \gls*{dmwf} estimation filter $\tW$ from~\eqref{eq:tW_partitioning}.
The network-wide version $\Wk$ of the \gls*{dmwf} filters $\tW$ at node $k$ can be written as:

\begin{align}\label{eq:nw_dmwf}
    \Wk = \begin{bmatrix}
        \Wk[k1]\\
        \vdots\\
        \Wkk\\
        \vdots\\
        \Wk[kK]
    \end{bmatrix}
    = \begin{bmatrix}
        \Pktq[1][k]\Gk[k1]\\
        \vdots\\
        \Wkk\\
        \vdots\\
        \Pktq[K][k]\Gk[kK]
    \end{bmatrix}
    =
    \begin{bmatrix}
        \Gam[1]\cCkqu[1][k]\Mku[11]\Gk[k1]\\
        \vdots\\
        \Wkk\\
        \vdots\\
        \Gam[K]\cCkqu[K][k]\Mku[KK]\Gk[kK]
    \end{bmatrix},
\end{align}

\noindent
where it can be verified that $\Wk^\Her\yk[]=\tW^\Her\ty$.

Since~\eqref{eq:estimation_lmmse_fc} is strictly convex, computing its solution (i.e., optimizing $\Wkk$ and $\{\Gk[kq]\}_{q\in\Kbq[k]}$ to estimate $\dk$) is sufficient to obtain the optimal solution $\Wk = \hWk\fa k\in\K$. Indeed, after the optimization~\eqref{eq:tW_dmwf_mwf}, the filters $\Wkk$ and $\{\Gk[kq]\}_{q\in\Kbq[k]}$ will adopt the optimal values:

\begin{align}
    \Gk[kq] &= \Mqtk[qq]^{-1}\Mkqu\fa q\in\Kbq[k],\\
    \Wkk&=\hWk[kk],
\end{align}

\noindent
that give $\Wk[kq] = \hWk[kq]\fa q\in\K$. This proves the optimality of the \gls*{dmwf}.\hfill$\blacksquare$

\bibliographystyle{IEEEbib_mod}
\bibliography{IEEEabrv,refs}

\end{document}